\documentclass[aps,prd,preprintnumbers,showpacs,showkeys,nofootinbib,
superscriptaddress,fleqn,floatfix,tightenlines,10pt]{revtex4-1}
\usepackage{amsmath,amsfonts,amssymb,amscd,amsxtra,amsthm}
\usepackage{graphicx}  
\usepackage{epstopdf}
\usepackage{dcolumn}  
\usepackage{bm}          
\usepackage{slashed}
\usepackage{cancel}
\usepackage{float}
\usepackage{mathtools} 
\usepackage{amsbsy}
\usepackage{amstext}

\usepackage[utf8]{inputenc} 
\usepackage{booktabs}
\usepackage[normalem]{ulem} 
\usepackage[dvipsnames]{xcolor} 
\usepackage{tabularx}
\usepackage{enumitem}  
\usepackage{array} 
\usepackage{slashed}
\usepackage{tikz}
\usepackage{float}
\usepackage{multirow}
\renewcommand\sout{\bgroup \color{red} \ULdepth=-.5ex \ULset}

\makeatletter

\begin{document}  
\preprint{INHA-NTG-12/2019}
\title{Pion mass dependence of the electromagnetic form factors of 
  singly heavy baryons}   
\author{June-Young Kim}
\email[E-mail: ]{Jun-Young.Kim@ruhr-uni-bochum.de}
\affiliation{Institut f\"ur Theoretische Physik II, Ruhr-Universit\"at
  Bochum, D-44780 Bochum, Germany}
\affiliation{Department of Physics, Inha University, Incheon 22212,
Republic of Korea} 
\author{Hyun-Chul Kim}
\email[E-mail: ]{hchkim@inha.ac.kr}
\affiliation{Department of Physics, Inha University, Incheon 22212,
Republic of Korea}
\affiliation{School of Physics, Korea Institute for Advanced Study 
  (KIAS), Seoul 02455, Republic of Korea}
\date{\today}
\begin{abstract}
  We study the electromagnetic form factors of the lowest-lying singly
heavy baryons with spin 1/2 within the framework of the
chiral quark-soliton model, focusing on the comparison
with recent lattice data.  To compare the present results
quantitatively with the lattice data, it is essential to treat the
pion mass as a variable parameter, i.e., to employ the unphysical
values of the pion mass, which are used in lattice
calculations. While the results with the physical value of the pion
mass fall off faster than those from the lattice calculations as the
momentum transfer increases, the extrapolated results with larger pion
masses get closer to the lattice data. This indicates that the pion
mean-field approach describes structures of both the light and singly
heavy baryons. 
\end{abstract}
\maketitle
\section{Introduction}
It is of utmost importance to study electromagnetic (EM) properties of
a baryon in understanding its structure. While the EM structures of
light baryons have been investigated well over decades, those of singly
heavy baryons have not been much examined. The reason is that
it is rather difficult to get access to EM properties of 
singly heavy baryons experimentally. On the other hand, very recently,
EM form factors of the singly heavy baryons~\cite{Can:2013tna} have
been investigated in a lattice QCD, which provide essential
information on the EM structure of them.  In Refs.~\cite{Can:2013tna}
large values of the unphysical pion mass were employed.
When one computes observables of hadrons, it is critical to consider
those values of the unphysical pion mass used in lattice calculations,
so that one can compare quantitatively 
the results from a certain model with those from the lattice data.

References~\cite{Goeke:2005fs, Goeke:2007fq} investigated the nucleon
mass and energy-momentum tensor form factors of the nucleon,
emphasizing the comparison of the results with the lattice data, based
on the chiral quark-soliton model ($\chi$QSM). In particular, 
Ref.~\cite{Goeke:2005fs} showed that the $\chi$QSM describes 
remarkably well the lattice data on the nucleon mass.
This $\chi$QSM~\cite{Diakonov:1987ty} was constructed based on
an idea that a baryon can be viewed as a state of $N_c$ (the number of
colors) valence quarks bound by the pion mean field. This mean-field
approach is justified in the large $N_{c}$ limit~\cite{Witten:1979kh,
  Witten:1983tx}, since the quantum fluctuation of the meson fields is
of order $1/N_c$, which can be neglected in this limit. The presence
of the $N_c$ valence quarks gives rise to the vacuum polarization that
produces the pion mean fields. Then the pion mean fields affect
self-consistently the $N_c$ valence quarks. This self-consistent
process makes a baryon emerge as a chiral soliton, which is a bound
state of the $N_c$ valence quarks. The $\chi$QSM has been successfully
used to explain properties of the SU(3) light
baryons~\cite{Wakamatsu:1990ud, Christov:1995vm, Diakonov:1997sj} (see
also Ref.~\cite{Alkofer:1994ph} that took a somewhat different 
approach). The $\chi$QSM was extended to a singly heavy baryon that
can be regarded as a bound state of $N_c-1$ valence quarks in the
large $N_c$ limit~\cite{Diakonov:2010tf, Yang:2016qdz}. A heavy quark
inside the singly heavy baryon can be treated as a static color source
when the heavy quark mass ($m_Q$) is taken to be infinitely heavy. 
The explicit effects of the heavy-quark mass only appear in the
splitting of the baryon sextet representations that are degenerate in
the limit of $m_Q\to\infty$.  
The model was successfully applied to the description of properties of
the lowest-lying heavy baryons such as the mass 
splittings~\cite{Yang:2016qdz, Kim:2018xlc, Kim:2019rcx}, isospin mass
differences~\cite{Yang:2020klp}, magnetic and transition magnetic
moments~\cite{Yang:2018uoj, Yang:2019tst}, and radiative
decays~\cite{Yang:2019tst}.

As already explained in Refs.~\cite{Goeke:2005fs,
  Goeke:2007fq}, an original purpose of studying the pion mass
dependence within the $\chi$QSM is to connect the results from 
chiral perturbation theory ($\chi$PT) and those from lattice QCD,
which is often called the chiral extrapolation. The $\chi$QSM serves
well for this purpose. Even though one takes a very large value of the
pion mass, the $\chi$QSM provides a stable chiral soliton. When one
takes a limit of the heavy pion mass, we find that a light quark tends
to behave as a heavy quark. Consequently the pion mean field seems to
be suppressed as the pion mass increases, which will be explicitly
shown later. On the other hand, the opposite limit, i.e., the chiral
limit, does not commute with the large $N_c$
limit~\cite{Gasser:1980sb, Dashen:1993jt}.  In the $\chi$QSM, we
adopt the following strategy: 
one first take the limit of $N_c\to\infty$ while keeping $m_\pi$
finite. Then, the $\chi$QSM produces properly a leading non-analytic
term of the nucleon mass expanded with respect to the pion
mass~\cite{Cohen:1992uy, Schuren:1991sc, Schweitzer:2003sb}. This 
indicates that the $\chi$QSM inheres a correct chiral behavior. This
is natural, since the model incorporates chiral symmetry and its
spontaneous breakdown.

Very recently, the electric monopole ($E0$) and magnetic dipole ($M1$)
form factors of the lowest-lying singly heavy baryons were
investigated within the framework of the 
$\chi$QSM~\cite{Kim:2018nqf}. In the present work, we extend the
previous work by extrapolating the experimental value of the physical
pion mass to unphysical ones that correspond to the values employed in
the lattice calculations. As mentioned previously, a virtue of the
$\chi$QSM is that it can be easily associated with a value of the
unphysical pion mass that is used in any lattice calculation. 
Thus, in the present work, we will
examine the pion mass dependence of the EM form factors of the singly
heavy baryons with spin 1/2 in the context of a recent lattice
work~\cite{Can:2013tna}. We will soon see that by incorporating the
unphysical values of the pion mass the present results describes
better the those from the lattice data.

The present paper is organized as follows: In Section II, we recapitulate
briefly how the EM form factors of the singly heavy baryons are
computed within the framework of the $\chi$QSM. In Section III, we
present the numerical results of the form factors in comparison with
the lattice data. In Section IV, we summarize the present work and
draw conclusions.   

\section{Electromagnetic form factors in the $\chi$QSM} 
Since we have presented the formalism as to how the EM form factors of
singly heavy baryons with spin 1/2 were derived in
Ref.~\cite{Kim:2018nqf}, we will briefly recapitulate it, emphasizing 
dependence of the EM form factors on the pion mass.
The EM current including a heavy quark is defined by
\begin{align}
  \label{eq:EMcurrent}
J_\mu(x) = \bar{\psi}(x) \gamma_\mu \hat{\mathcal{Q}} \psi(x) + e_Q
  \bar{\Psi}(x) \gamma_\mu \Psi(x),   
\end{align}
where the first term of Eq.~\eqref{eq:EMcurrent} denotes the EM
current of the light quarks whereas the second one corresponds to that
of the heavy quark. $\hat{Q}$ is the charge matrix of the light quarks
given by
\begin{align}
  \label{eq:chargeL}
\hat{\mathcal{Q}} =
  \begin{pmatrix}
    \frac23 & 0& 0    \\ 0 & -\frac13 & 0
    \\ 0 & 0 & -\frac13 
  \end{pmatrix}
               = \frac12 \left(\lambda_3 + \frac1{\sqrt{3}} \lambda_8
               \right),
\end{align}
where $\lambda_3$ and $\lambda_8$ designate the flavor SU(3) Gell-Mann
matrices. $e_Q$ in the second part of Eq.~\eqref{eq:EMcurrent} is 
the corresponding charge of a heavy quark:  $e_c= 2/3$ for a charm
quark or $e_b= -1/3$ for a beauty quark. In the present pion
mean-field approach, we take the limit of the infinitely heavy-quark
mass ($m_Q\to \infty$), so that the second part of
Eq. ~\eqref{eq:EMcurrent} provides only the constant charge to the
electric form factor of a singly heavy baryon. Since the magnetic form
factor of a heavy quark is proportional to its inverse mass, i.e.,
$\bm{\mu} \sim (e_Q/m_Q) \bm{\sigma}$, we can safely neglect the
heavy-quark contribution to the magnetic form factor.  

The EM form factors of the singly heavy baryons    
are related to the matrix element of the EM current between
the singly heavy baryon states with spin 1/2 as 
\begin{align}
\langle B,\,p' | J_\mu(0) |B, \,p\rangle = \overline{u}_{B}(p',\,\lambda') 
  \left[\gamma_\mu F_1(q^2) + i\sigma_{\mu\nu} \frac{q^\nu}{2M_B}
  F_2(q^2)\right] u_B(p,\,\lambda),   
\label{eq:MatrixEl1}
\end{align}
where $q^2$ denotes the square of the four-momentum transfer
$q^2=-Q^2$ with $Q^2 >0$. $u_B(p,\,\lambda)$ stands for the Dirac
spinor with four-momentum $p$ and the helicity $\lambda$ for a baryon
$B$ with spin 1/2. The EM Sachs form factors $G_E(Q^2)$ and $G_M(Q^2)$
can be expressed in terms of the Dirac and Pauli form factors
$F_1(Q^2)$ and $F_2(Q^2)$  
\begin{align}
G_E^B(Q^2) &= F_1^B (Q^2) - \tau F_2^B (Q^2), \cr
G_M^B(Q^2) &= F_1^B (Q^2) + F_2^B(Q^2),
\end{align}
with $\tau=Q^2/4M_B^2$. In the Breit frame, the matrix elements for
the temporal and spatial components of the EM current give the
electric and magnetic form factors, respectively. 
\begin{align}
 \langle B,\,p' |J_0(0)
              |B,\,p\rangle &= G_E^B(Q^2) \delta_{\lambda'\lambda}, \cr 
\langle B,\,p'
             |J^k(0) |B,\,p\rangle &=  \frac{i}{2M_{B}} (\bm{\sigma}
                                     \times
                                     \bm{q})^{k}_{\lambda'\lambda}
                                     G_M^B(Q^2),   
\end{align}
where the subscripts $\lambda'$ and $\lambda$ indicate the matrix
elements in the two-component helicity basis. 
Thus, we can evaluate the EM form factors of the singly heavy baryons
by computing the matrix elements of the EM current within the
framework of the $\chi$QSM. 

The $\chi$QSM is described by the low-energy effective QCD partition
function in Euclidean space
\begin{align}
\label{eq:partftn}
\mathcal{Z}_{\chi\mathrm{QSM}} = \int \mathcal{D} U \exp
  (-S_{\mathrm{eff}}),   
\end{align}
where the quark fields have been integrated out. $S_{\mathrm{eff}} $
denotes the effective chiral action 
\begin{align}
S_{\mathrm{eff}}[U] \;=\; -N_{c}\mathrm{Tr}\ln (i\rlap{/}{\partial} +
  i M U^{\gamma_{5}} + i \hat{m} )\, ,
\label{eq:echl}
\end{align}
with the number of colors, $N_{c}$. Here, $M$ stands for the dynamical
quark mass that is the only free parameter of the model. We will
discuss later the procedure of fixing parameters including $M$. 
$U^{\gamma_{5}}$ represents the chiral field
\begin{align}
  \label{eq:4}
U^{\gamma_5} = \exp(i\pi^a \lambda^a \gamma_5) = \frac{1+\gamma_5}{2}
  U + \frac{1-\gamma_5}{2} U^\dagger,
\end{align}
with $U=\exp(i\pi^a \lambda^a)$. $\pi^a$ designates the
pseudo-Nambu-Goldstone fields with the flavor index $a$ running over
$a=1,\cdots 8$. $\hat{m}$ is the matrix of the current-quark masses
$\hat{m} = \mathrm{diag}(m_{\mathrm{u}},\,m_{\mathrm{d}},\,m_{\mathrm{s}})$.
We will assume isospin symmetry in the present work, so that 
$m_{\mathrm{u}}=m_{\mathrm{d}}$. The average mass of the up and down
quarks will be defined by $m_0 =
(m_{\mathrm{u}}+m_{\mathrm{d}})/2$. The effective chiral action can be
expressed in terms of the Dirac one-body Hamiltonian $h(U)$
\begin{align}
  \label{eq:1}
S_{\mathrm{eff}} = -N_c \mathrm{Tr}\ln \left(\partial_4 + h(U) + \gamma_4
 \hat{m} - \gamma_4 m_0 \bm{1}\right),  
\end{align}
where $h(U)$ is written by
\begin{align}
  \label{eq:3}
h(U) = -i \gamma_4\gamma_i \partial_i + \gamma_4 M U + \gamma_4 m_0
  \bm{1}.   
\end{align}
We introduce a new mass matrix for the current quarks 
\begin{align}
\delta m  = \hat{m} - m_0\bm{1} = \frac{-m_{0} + m_{s}}{3}\bm{1} +
\frac{m_{0} - m_{s}}{\sqrt{3}}  \lambda^{8} =
m_{1} \bm{1} + m_{8}  \lambda^{8}\,,
\label{eq:deltam}
\end{align}
where $m_1$ and $m_8$ are respectively defined by 
\begin{align}
m_1=\frac13(-m_{0} +m_{\mathrm{s}}), \;\;\; 
m_8=\frac1{\sqrt{3}} (m_{0} -m_{\mathrm{s}}).
\end{align}

The integral over the $U$ field can be performed by the saddle-point
approximation that is justified in the large $N_c$ limit. Since we
have to preserve the hedgehog symmetry given by
\begin{align}
\pi^a = n^a P(r),\;\;\; \pi^b=0   \mbox{ with } b=4,\cdots, 8  
\end{align}
with the profile function $P(r)$ of the classical soliton, we need to
embed the SU(2) $U_0$ field into SU(3)~\cite{Witten:1983tx}
\begin{align}
  \label{eq:6}
U =
  \begin{pmatrix}
    U_0 & 0 \\ 0 & 1
  \end{pmatrix} ,
\end{align}
where $U_0$ denotes the SU(2) chiral field
\begin{align}
  \label{eq:7}
U_0 = \exp[ i n^a   \tau^a P(r) ].
\end{align}

As shown explicitly in Ref.~\cite{Kim:2018xlc, Kim:2019rcx}, the
classical mass of a singly heavy baryon can be derived by computing
the baryon correlation function in large Euclidean time. Then, the
classical soliton mass is obtained to be the sum of the energies of
the valence and sea quarks, $M_{\mathrm{sol}}=(N_c-1)
E_{\mathrm{val}}+E_{\mathrm{sea}}$. Then, the classical equation of
motion can be derived by minimizing the energy of the classical
soliton   
\begin{align}
\left.\frac{\delta}{\delta P(r)}[ (N_c-1)
  E_{\mathrm{val}} + 
  E_{\mathrm{sea}}]\right|_{P_c} = 0, 
\label{eq:saddle}
\end{align}
where $P_c$ is the profile function of the soliton at the stationary
point, which is just a solution of the pion mean fields.
Hence, the soliton mass for the singly heavy baryon is finally
obtained as  
\begin{align}
M_{\mathrm{sol}} = (N_c-1) E_{\mathrm{val}}(P_c) +
  E_{\mathrm{sea}}(P_c).    
\label{eq:solnc}
\end{align}
The classical mass $M_{\mathrm{cl}}$ is defined as
\begin{align}
M_{\mathrm{cl}} = M_{\mathrm{sol}} + m_{Q},
\end{align}
where $m_{Q}$ is the effective heavy quark mass which includes the
binding energy of the heavy quark.

Since we are interested in computing the EM form factors of the singly
heavy baryons with the pion mass varied, we have to derive the profile
function, given a value of the unphysical pion mass (see
Appendix~\ref{app:A} for details as to how we can fix the parameters
in the mesonic sector). Consequently, the soliton mass for the singly
heavy baryon depends on the pion mass. If the value of the pion mass
or that of $m_0$ grows, the soliton mass will converge on $2m_0$,
i.e.,  
\begin{align}
  \label{eq:8}
\lim_{m_0\to \infty}  M_{\mathrm{sol}} (m_0) = (N_c-1) m_0, 
\end{align}
which was already shown in Ref.~\cite{Goeke:2005fs}. We will call it
the relation for the soliton mass in the limit of large light quark
mass. 
\begin{figure}[ht]
  \centering
  \includegraphics[scale=0.28]{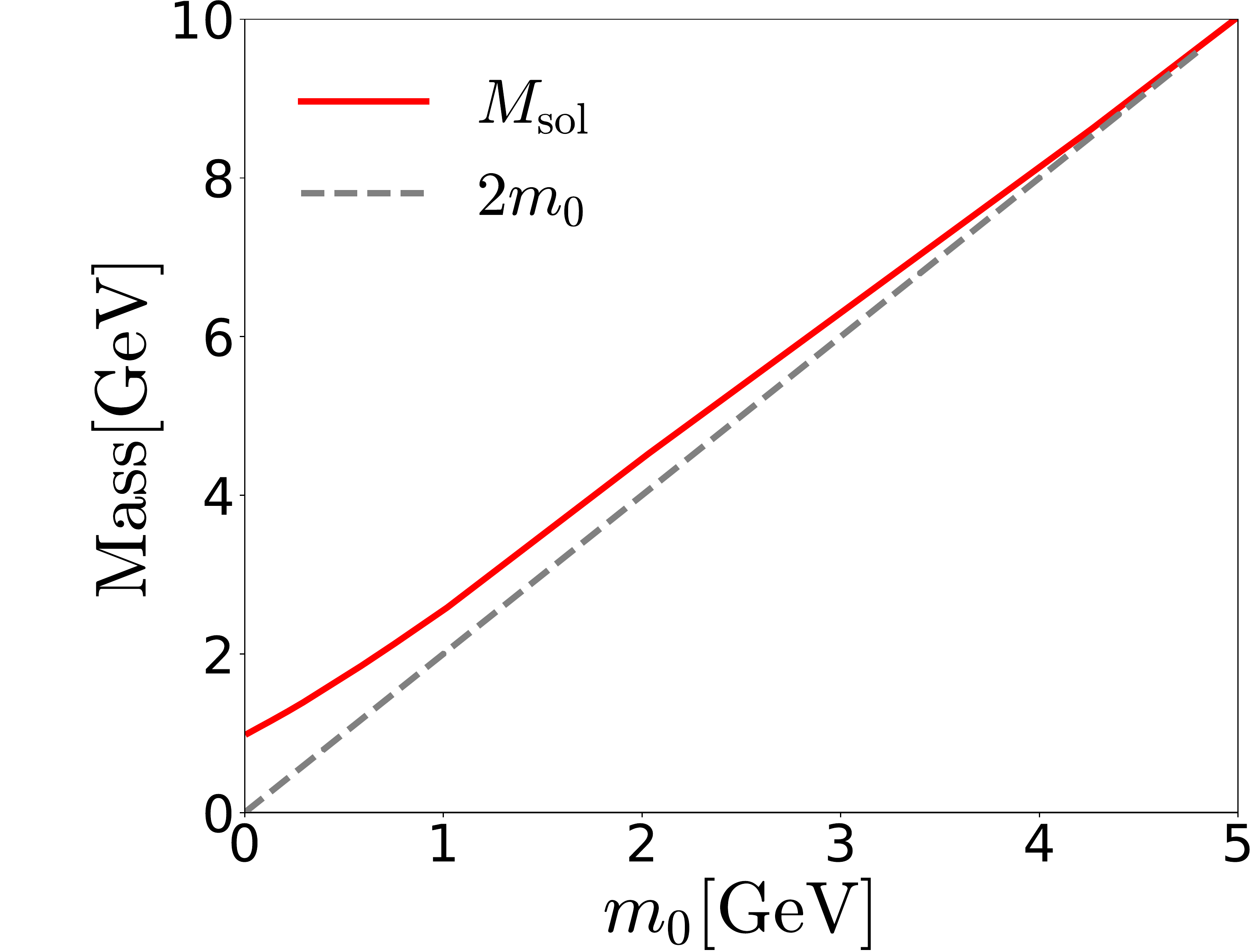}
  \caption{Dependence of the soliton mass on $m_0$. The solid curve
    draws the result for the soliton mass as the $m_0$ varied. the
    dashed one depicts $2m_0$.}
    \label{fig:1}
  \end{figure}
  In Fig.~\ref{fig:1} we draw the soliton mass as a function of
  $m_0$. The result indicates that the soliton mass converges on $2m_{0}$
as $m_0$ increases. The numerical result indeed satisfies
Eq.~\eqref{eq:8}. It means that as $m_0$ increases, the effects of the
pion mean field are relatively reduced. 

The general formalism for the EM form factors of the singly heavy
baryons with spin 1/2 was already given in Ref.~\cite{Kim:2018nqf} in
detail. Thus, we will only compile the final expressions in the
following:  
\begin{align}
{{G}}^{B}_{E}  (q^{2})&= \int d^{3} z j_{0}(|\bm{q}| |\bm{z}|)
  {\mathcal{G}}^{B}_{E} (\bm{z}) + G_E^Q(q^2), \\
{{G}}^{B}_{M}  (q^{2})&=\frac{ M_{B}}{|\bm{q}|} \int d^{3} z
  \frac{j_{1}(|\bm{q}| |\bm{z}|)}{ |\bm{z}|} {\cal{G}}^{B}_{M}
  (\bm{z}) , 
\label{eq:app1}
\end{align}
where 
\begin{align}
{\cal{G}}^{B}_{E} (\bm{z})=& \frac{1}{\sqrt{3}} \langle
  D^{(8)}_{Q8}\rangle_B \mathcal{B}(\bm{z}) -
  \frac{2}{I_{1}} \langle
  D^{(8)}_{Qi} \hat{J}_{i} \rangle_B
  {\cal{I}}_{1}(\bm{z}) -
  \frac{2}{I_{2}} \langle
  D^{(8)}_{Qp} \hat{J}_{p} \rangle_B
  {\cal{I}}_{2}(\bm{z}) \cr 
    & -\frac{4 m_{8}}{I_{1}} \langle D^{(8)}_{8i} D^{(8)}_{Qi}
      \rangle_B (I_{1}{\cal{K}}_{1}(z) - K_{1}{\cal{I}}_{1}(z)) 
  -\frac{4 m_{8}}{I_{2}} \langle D^{(8)}_{8p} D^{(8)}_{Qp}
   \rangle_B (I_{2}{\cal{K}}_{2}(z) - K_{2}{\cal{I}}_{2}(z)) \cr 
 &-2\left (  \frac{m_1}{\sqrt{3}}\langle D^{(8)}_{Q8} \rangle_B +
   \frac{m_8}{3}\langle D^{(8)}_{8 8}D^{(8)}_{Q8} \rangle_B  \right)
   {\cal{C}}(\bm{z}),
     \label{eq:9}
\end{align}
\begin{align}
   {\cal{G}}^{B}_{M}(\bm{z}) &=  \langle D^{(8)}_{Q3 }
  \rangle_B  \left(  {\cal{Q}}_{0} (\bm{z})  + \frac{1}{I_{1}}
 {\cal{Q}}_{1} (\bm{z}) \right) -  \frac{1}{\sqrt{3}} \langle
 D^{(8)}_{Q 8}\hat{J}_{3} \rangle_B \frac{1}{I_{1}}
 {\cal{X}}_{1} (\bm{z}) - \langle
 d_{pq3} D^{(8)}_{Qp} \hat{J}_{q} \rangle_B
 \frac{1}{I_{2}}  {\cal{X}}_{2} (\bm{z})   \cr 
& + \frac{2}{\sqrt{3}} m_{8} \langle D^{(8)}_{83}
  D^{(8)}_{Q8} \rangle_B \left(\frac{K_{1}}{I_{1}}{\cal{X}}_{1}
  (\bm{z}) -   {\cal{M}}_{1} (\bm{z})\right)  
  +2 m_{8} \langle  d_{pq3}  D^{(8)}_{8p}
  D^{(8)}_{Qq} \rangle_B \left(\frac{K_{2}}{I_{2}}{\cal{X}}_{2} (\bm{z})
     -    {\cal{M}}_{2} (\bm{z})\right)  \cr 
& - 2  \left( m_{1} \langle D^{(8)}_{Q3} \rangle_B +
  \frac{1}{\sqrt{3}} m_{8} \langle D^{(8)}_{88} D^{(8)}_{Q3}
  \rangle_B  \right) {\cal{M}}_{0} (\bm{z}).
\label{eq:final}
\end{align}
The explicit expressions for those densities, and moments of inertia
$I_{1,2}$ and $K_{1,2}$ are given already in
Ref.~\cite{Kim:2018nqf}. $G_E^Q(q^2)$ in Eq.~\eqref{eq:app1}
represents the heavy-quark contribution to an electric form factor of
a singly heavy baryon. In the limit of $m_Q\to\infty$, it gives just
the charge of the corresponding heavy quark.

\section{Results and discussion}
Before we present the numerical results, we first explain briefly
how to determine the model parameters. We fix them first in the
mesonic sector. Since the pion decay constant diverges
logarithmically, which arises from the corresponding quark loop, we
need to introduce a regularization scheme. In the present work, we
adopt the proper-time regularization with the cutoff mass $\Lambda$
that can be fixed by reproducing the experimental value of the pion
decay constant $f_{\pi} = 93$~MeV. The average value of the up and
down current quark masses, $m_0$, is determined by reproducing the
physical value of the pion mass $m_\pi=140$ MeV. The only free
parameter is then the dynamical quark mass, $M$, which will be
determined by reproducing various properties of the proton. The best
value turns out to be $M=420$~MeV and we keep using this value also
for the heavy baryon sector. 

Since we want to extrapolate the present model by employing various
different values of the unphysical pion mass, we have to proceed to
fix the parameters very carefully. As we explain in
Appendix~\ref{app:A} in detail, one should distinguish $M$ from 
$M'=M+m_0$ that appears in the expressions for the quark condensate
and pion decay constant. The value of the dynamical quark mass $M$ is
always fixed to be $420$ MeV. We want to mention that there is one
caveat related to the pion decay constant. In effect, the value of the
pion decay constant increases as that of the unphysical pion mass
increases in lattice calculations. However, since the pion decay
constant is divergent logarithmically, its change is rather mild as
the pion mass varies. Indeed, the value of the pion decay constant
from the lattice QCD~\cite{Noaki:2008iy, Durr:2013goa} is enhanced by 
about 30~\% when the value of $m_{\pi}$ is taken to be approximately
0.5 GeV.  This means that it is still approximately valid to keep
using the experimental value of the pion decay constant to fix the
cutoff mass $\Lambda$. Thus, we will continue to use it to fix the
cutoff mass as our prescription. On the other hand, the average mass
of the up and down valence quarks $m_0$ depends directly on the value
of the unphysical pion mass, which we have to consider seriously.

This strategy for comparison with the lattice results was already
discussed in Ref.~\cite{Goeke:2005fs} in detail. Of course we could
have taken the values of $f_\pi(m_\pi)$ produced in lattice
calculations as input. This means that both the pion
decay constant and the quark condensate securely increase as $m_\pi$
increases. In this case, the results for the EM form factors of singly
heavy baryons are obtained to be almost the same as the present
ones. However, there is a caveat in this analysis. If one 
increases the pion mass larger than 400 MeV, then the soliton solution
does not exist. This is no wonder: the parameters $f_\pi$, $m_\pi$,
$m_0$, and $\Lambda$ in the present model are interrelated, so that we
are not able to change one of them independently while keeping the
soliton solution stable. Thus, we will rather regard the discrepancy
for the pion decay constant arising from the comparison with the
lattice results as the model accuracy, since the present model is
used to describe the observables within the $(5-30)\,\%$
accuracy.

Given a value of the unphysical pion mass, then we are able to fix
$\Lambda$ for regulators and the current quark 
mass $m_{0}$ by using Eq.~\eqref{eq:pi} and
Eq.~\eqref{eq:pionmass}. From 
those fixed parameters, we get the chiral condensate, also known
as the chiral order parameter, $\langle \overline{\psi}\psi \rangle$
defined in Eq.~\eqref{eq:gap}. It characterizes the strength of the 
spontaneous breakdown of chiral symmetry. As shown in
Eq.~\eqref{eq:gap}, it is inevitable to provide the numerical value of
$M$ to determine the quark condensate. The same is true also for the
pion decay constant (see Eq.~\eqref{eq:pi}). Another physical
implication of the dynamical quark mass is the coupling strength
between the quark and pNG fields. The strange current quark mass is
taken to be $m_{s} = 180$~MeV to reproduce the mass splitting of
flavor SU(3) light baryons~\cite{Blotz:1992pw} and singly heavy
baryons~\cite{Kim:2018xlc}.  

Now, a natural question may arise. Is the dynamics of the $\chi$QSM 
appropriate for extrapolating the pion mass to the unphysical ones?
We can answer this question as follows: Firstly, the effective chiral
action given in Eq. ~\eqref{eq:echl} can be derived from the QCD
instanton vacuum~\cite{Diakonov:1985eg, Diakonov:1997sj}, which may be
considered as a low-energy effective model of QCD. Actually, the
dynamical quark mass from the instanton vacuum depends on the quark
momentum. This momentum-dependent quark mass also plays a role of a
regulator. However, we turn off the momentum dependence of the
dynamical quark mass to avoid theoretical complexities and introduce
an explicit regularization scheme to tame the divergences arising from
the quark loops. Secondly, since the effective chiral action
complies with chiral symmetry and its spontaneous breakdown, it
naturally contains all orders of the effective chiral Lagrangians in
the leading order of $N_c$. This can be shown explicitly by the
derivative expansion~\cite{Diakonov:1987ty, RuizArriola:1991gc,
  Choi:2003cz}. Thus, the $\chi$QSM respects at least important
symmetries and properties of low-energy QCD, so that it is in a proper
position to be confronted with lattice QCD.  

\setlength{\tabcolsep}{5pt}
\renewcommand{\arraystretch}{1.5}
\begin{table}[htp]
\centering
\caption{Dependence of the valence- and sea-qaurk energies, and the
  soliton mass on the values of the pion mass.}  
\label{tab:1}
\begin{tabular}{ c | c c c c c c } 
\hline 
  \hline 
$m_{\pi} [\mathrm{MeV}]$ & $m_{0} [\mathrm{MeV}]$  & $\Lambda
 [\mathrm{MeV}]$ &   $-\langle \overline{\psi}\psi \rangle^{-1/3}
  [\mathrm{MeV}]$  & $E_{\mathrm{val}}
 [\mathrm{MeV}]$   &  $E_{\mathrm{sea}}   [\mathrm{MeV}]$ 
  & $M_{\mathrm{sol}} [\mathrm{MeV}]$\\   
\hline 
$140$ & 18 & 637 & 210 & 645 & 354 & 999 \\
$300$ & 75 & 645 & 206 & 717 & 362 & 1078 \\
$410$ & 130 & 659 & 205 & 786 & 366 & 1152 \\
$570$ & 219 & 689 & 204 & 908 & 370 & 1278 \\
$700$ & 295 & 718 & 204 & 1019 & 371 & 1380 \\
\hline
\hline
\end{tabular}
\end{table}
In Table.~\ref{tab:1}, we list the numerical values of the valence-
and sea-quark energies, and the soliton mass. As the pion mass
increases, both the valence- and sea-quark energies increase. In
consequence, the soliton mass also grows larger as a function of
$m_{\pi}$. As discussed in Ref.~\cite{Schweitzer:2003sb}, these results
for the nucleon mass as a function of $m_\pi^2$ are in good agreement
with the lattice data. 

\begin{figure}[htp]
\centering
\includegraphics[scale=0.28]{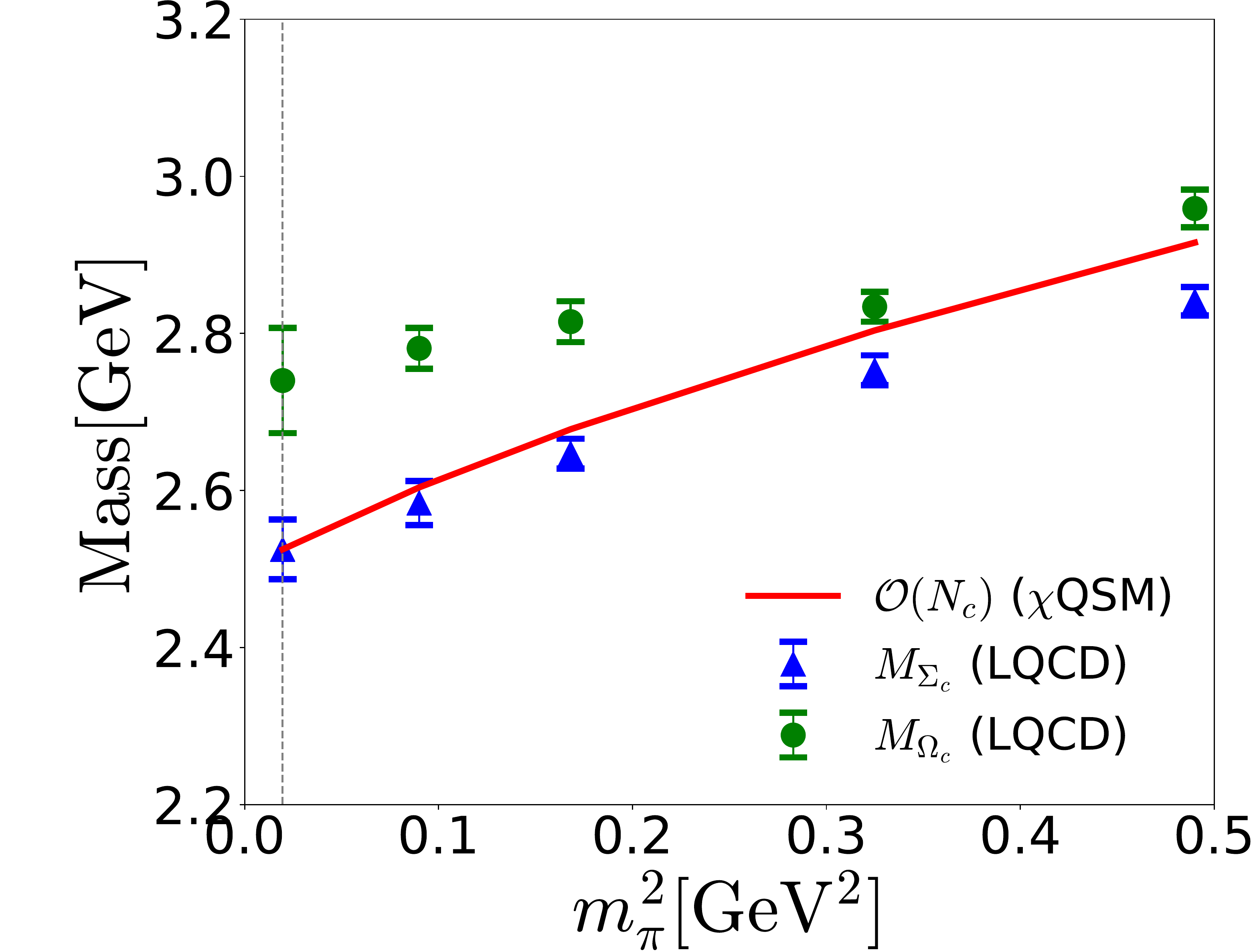}
\includegraphics[scale=0.28]{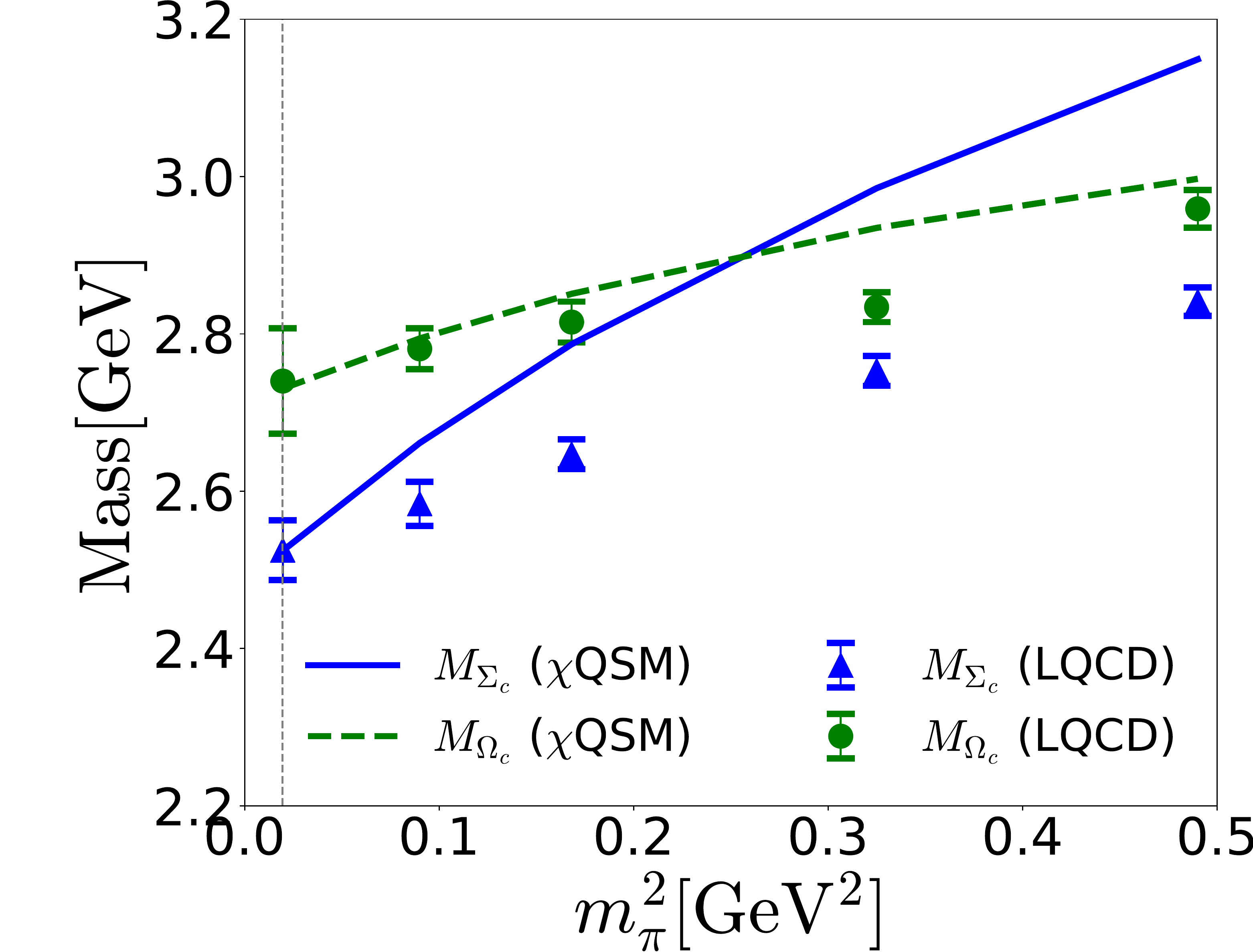}
\caption{Dependence of the masses of the singly heavy baryons,
  $\Sigma_c$ and $\Omega_c$. In the left panel, we draw the classical 
  mass as a function of $m_\pi^2$ whereas in the right panel, we
  depict the masses of $\Sigma_c$ and $\Omega_c$ as functions of
  $m_\pi^2$ in the solid and dashed curves, respectively. To compare
  these results with the lattice data, we normalize the classical
  mass by the lattice value of $M_{\Sigma_c}$ 
  at the physical pion mass, which is drawn as the vertical short
  dashed line. The lattice data are taken from Ref.~\cite{Can:2013tna}.}    
\label{fig:2}
\end{figure}
We now examine the dependence of the masses of
$\Sigma_c$ and $\Omega_c$ on the pion mass, which belong to the baryon
sextet with spin 1/2, comparing the present results with those from
lattice QCD. In the left panel of Fig.~\ref{fig:2}, we show the
numerical results for the classical mass $M_{\mathrm{cl}}$ 
as a function of $m_\pi^2$. Note that for comparison with the lattice
data we normalize the classical mass by the lattice value of the
$\Sigma_c$ mass at the physical value of the pion mass, $m_\pi=140$
MeV.  Interestingly, the result of the classical mass is in very good
agreement with the lattice data. In fact, the nucleon mass from the
$\chi$QSM was shown to be almost the same as the lattice data as
described in Ref.~\cite{Goeke:2005fs}. We want to mention that in
Ref.~\cite{Goeke:2005fs}, the nucleon mass was in fact the classical
mass. As mentioned in Introduction, the chiral limit ($m_\pi\to 0$)
does not commute with the large $N_c$ limit. In the $\chi$QSM, the
strategy is that one first take the limit of $N_c\to\infty$ while
keeping $m_\pi$ finite. Next, we take the chiral limit. In this case, 
the leading non-analytic term of the nucleon 
mass, which appears when the nucleon mass is expanded with respect to
the pion mass, is yielded to be
\begin{align}
  \label{eq:2}
M_N(m_\pi)^{\mathcal{O}(m_\pi^3)}  = k \frac{3 g_A^2}{32\pi f_\pi^2} m_\pi^3 .
\end{align}
This expression reproduces that obtained in one-loop $\chi$PT except
for the overall factor $k$. Chiral solitonic models give $k=3$ whereas
$\chi$PT provides $k=1$~\cite{Jenkins:1990jv}. Here, $g_A$ denotes the
axial charge of the nucleon. This has a very important physical
implication. In chiral solitonic models, the masses of the $\Delta$
isobar and the nucleon become degenerate in the large $N_c$ limit.
Moreover, taking the large $N_c$ limit with $m_\pi$ kept finite, we
find that $M_\Delta -M_N$ turns out to be much smaller than the pion
mass. This means that the $\Delta$ isobar must be considered as an
intermediate state in chiral loops, which provides as twice as the
nucleon contribution because of the different Clebsch-Gordan
coefficients, so we have $k=3$~\cite{Cohen:1992uy}. In contrast,
one-loop conventional  
$\chi$PT~\cite{Jenkins:1990jv} takes the opposite ordering, which
means that the chiral limit is taken first and then the large $N_c$
limit is considered in $\chi$PT. This indicates that the mass
difference $M_\Delta -M_N$ is much larger than $m_\pi$. So, the
contribution of the $\Delta$ isobar in the chiral loops is ignored,
which brings about the different value of $k$. 
As discussed in Ref.~\cite{Cohen:1992uy}, the ratio $d=(M_\Delta -
M_N)/m_\pi$ becomes 
infinity in conventional $\chi$PT whereas it goes to zero in chiral
solitonic approaches. However, the truth lies between these two
values.

Based on this argument, we can consider the degenerate masses of the
baryon sextet in the large $N_c$ limit. Hence, the left
panel of Fig.~\ref{fig:2} describes the representative mass of the
low-lying singly heavy baryons. In this sense, the result shown in
Fig.~\ref{fig:2} is indeed remarkable, since it describes both the
lattice data on the $\Sigma_c$ and $\Omega_c$ masses. Thus, the
$\chi$QSM provides a reliable framework for comparison of any
observables for the singly heavy baryons with the corresponding
lattice data.    
In the right panel of Fig.~\ref{fig:2}, we take a more realistic
position. So, we introduce the rotational $1/N_c$ and linear $m_s$
corrections, which also depend on the pion mass. While the present
result for the $\Sigma_c$ mass, which is depicted in the solid curve,
rises faster than the lattice data, that for the $\Omega_c$ mass,
drawn in the dashed curve, is in good agreement with the lattice
data. In fact, the mass spectra of the low-lying singly heavy baryons
were studied in Ref.~\cite{Kim:2018xlc}. The masses of the $\Sigma_c$
and $\Omega_c$ are expressed as
\begin{align}
  \label{eq:10}
M_{\Sigma_c} = M_{\bm{6}} + \frac23 \delta_{\bm{6}},\;\;\;
  M_{\Omega_c} = M_{\bm{6}} - \frac43 \delta_{\bm{6}}, 
\end{align}
where definitions of the parameters $M_{\bm{6}}$ and $\delta_{\bm{6}}$
can be found in Ref.~\cite{Kim:2018xlc}. The parameter
$\delta_{\bm{6}}$ is related to the linear $(m_{\mathrm{s}}-m_{0})$
corrections, so that it gives rise to the mass splitting in the baryon
sextet. Note that $\delta_{\bm{6}}$ has a negative value in the range
of $0\le m_\pi^2 \le 0.25\,\mathrm{Gev}^2$. Then, it is changed to
positive. This explains why the mass of $\Sigma_c$ is raised faster
than that of $\Omega_c$ as shown in the right panel of
Fig.~\ref{fig:2}. Note that the masses of $\Sigma_c$ and $\Omega_c$
coincide with each other at around $m_\pi^2 =
0.25\,\mathrm{GeV}^2$, where the average mass of the up and down
current quarks turns out to be the same as that of the strange current
quark. So, flavor SU(3) symmetry is restored at this point.    

As mentioned previously, to examine the pion mass dependence of the EM
form factors, we first have to compute the profile functions of the
chiral soliton given a value of the pion mass. To do that, we choose
its five different values: $m_\pi=140$ MeV (physical one), $m_\pi=300$
MeV, $m_\pi=410$ MeV, $m_\pi=570$ MeV, and $m_\pi=700$ MeV and
derive the new profile functions corresponding to these values of the pion
mass. Except for the physical one, all the values were employed by the
lattice calculation~\cite{Can:2013tna}.
So far, there is no experimental data on the EM form factors of the
singly heavy baryons. Thus, in the present work, we will
carefully compare the present results with those from a recent lattice
work~\cite{Can:2013tna}, considering the pion mass as a variable
parameter. 

\begin{figure}[htp]
\centering
\includegraphics[scale=0.28]{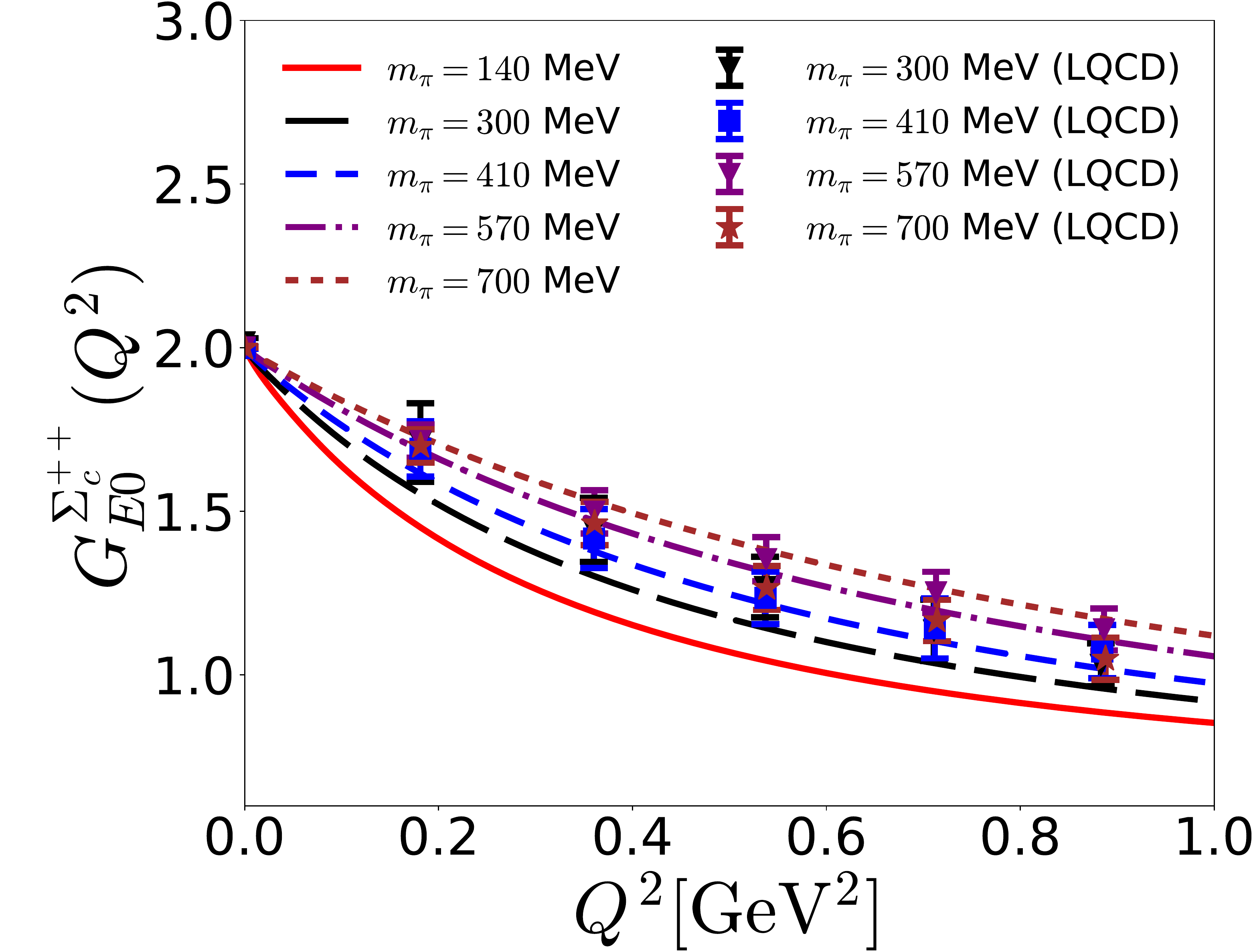}
\includegraphics[scale=0.28]{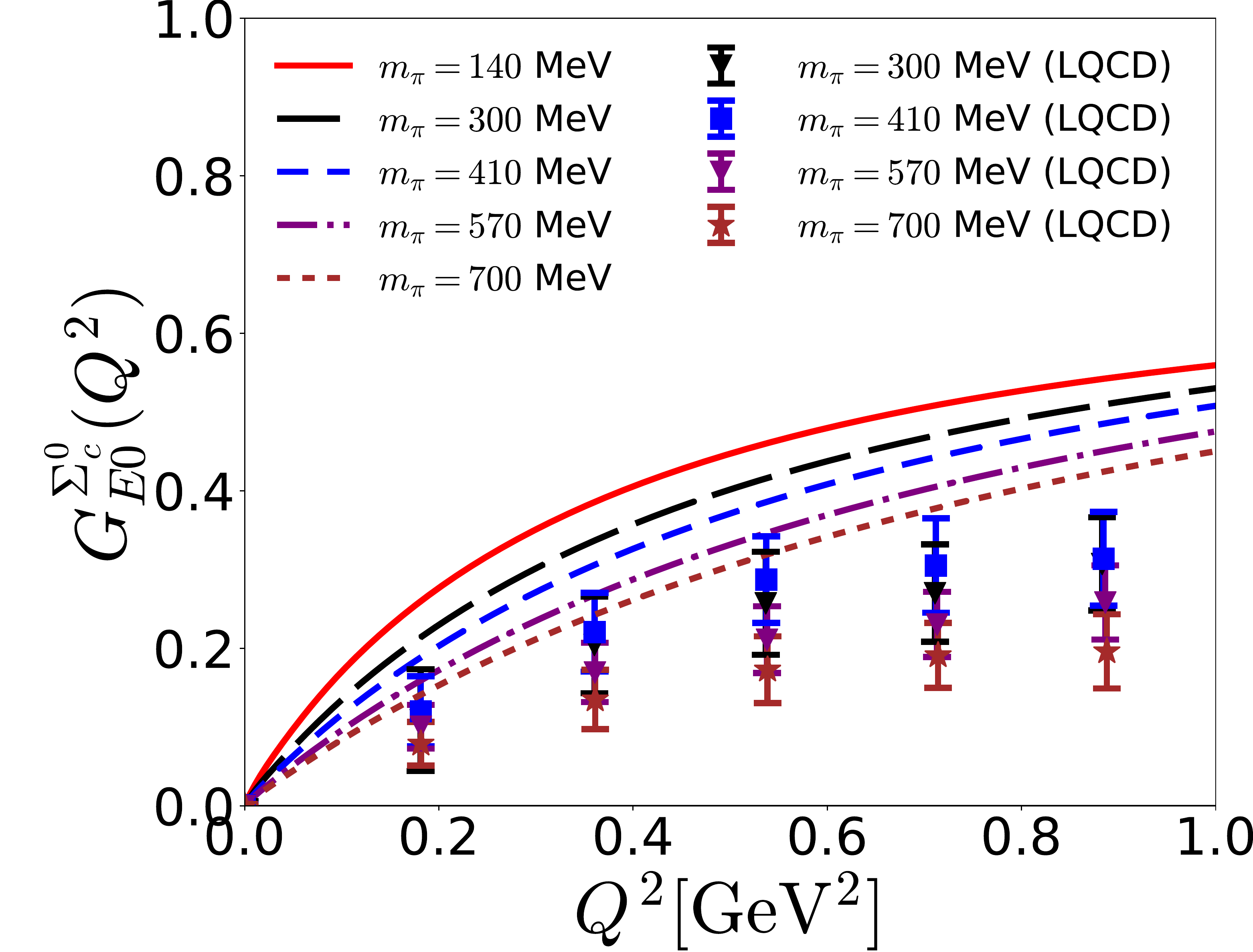}
\includegraphics[scale=0.28]{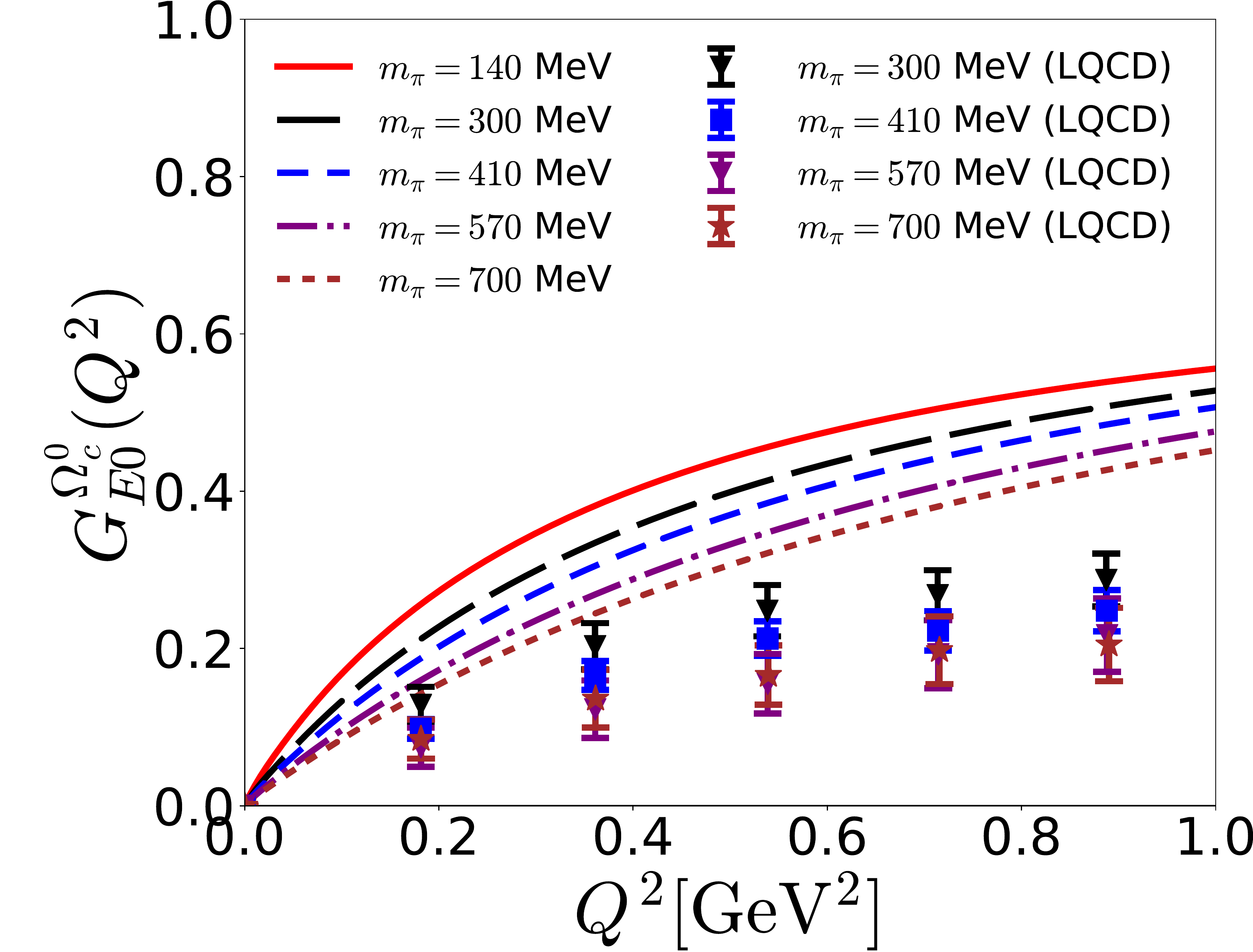}
\caption{Electric monopole form factors of the baryon sextet with
  spin-1/2 in comparison with the data from the lattice QCD. The data
  of the lattice QCD are taken from Ref.~\cite{Can:2013tna}. Note that
  the lattice data for the zero-charged electric form factors of the
  heavy baryons are taken from the private communication with
  U. Can~\cite{Can:2019}.}  
\label{fig:3}
\end{figure}
We first compare the results of the $E0$ form factors obtained from
the present model with those from the lattice
calculations~\cite{Can:2013tna}, though a part of  
the work with the physical pion mass was already done in
Ref.~\cite{Kim:2018nqf}. In Fig.~\ref{fig:3} we draw the electric form
factors of the $\Sigma_c^{++}$, $\Sigma_c^{0}$, and $\Omega_c^0$
baryons with spin 1/2 in comparison with the corresponding lattice data. we
extrapolate the physical pion mass $m_\pi=140$ MeV to the
unphysical ones of which the values are taken from those used in the
lattice calculation, i.e. four different values $m_\pi=300$ MeV, 410 MeV, 570
MeV, and 700 MeV. As expected, when we increase the values of the pion
mass, the results of the electric form factors fall off more slowly as
$Q^2$ increases. This is a well-known feature of the lattice
results. Thus, when one wants to compare results of any form factors
with those from lattice works, it is better to employ larger pion
masses that match the corresponding values used in the lattice
calculation. When the pion mass gets larger, the $Q^2$ dependences of
the electric form factors of the neutral heavy baryons increase more
slowly. This can be understood by examining the behavior of the
soliton profile function. As the pion mass increases larger than 140
MeV, the Yukawa tail of the soliton falls off faster than this
physical case. This indicates that the size of the baryon becomes more
compact than the physical one. Consequently, the results for the
electric form factors fall off more slowly. 
The numerical results for the $\Sigma_c^{++}$ electric form
factor are in agreement with the lattice data. Those for 
$\Sigma_c^0$ and $\Omega_c^0$ get closer to the data as the pion mass
increases.  

\begin{figure}[htp]
\centering
\includegraphics[scale=0.28]{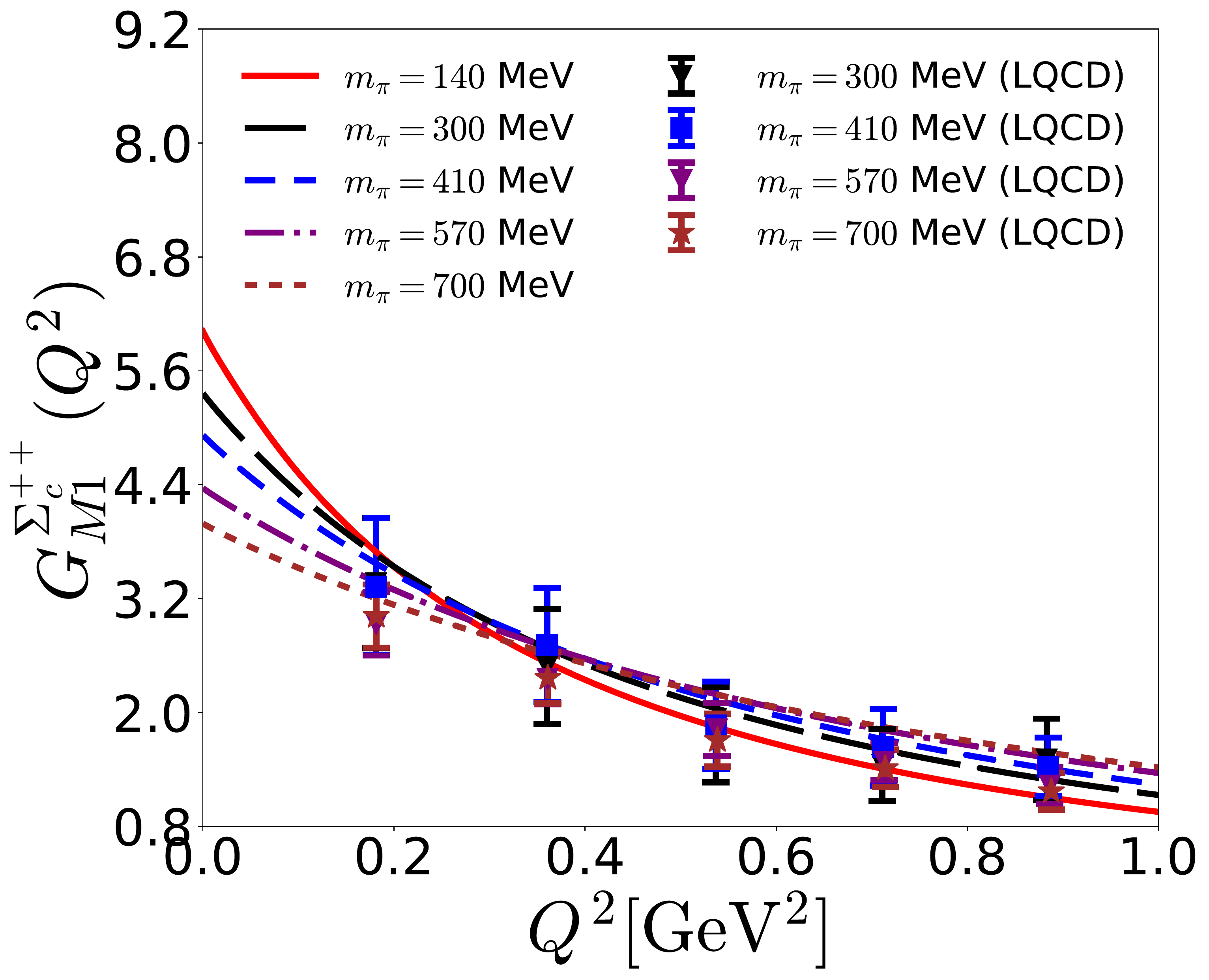}
\includegraphics[scale=0.28]{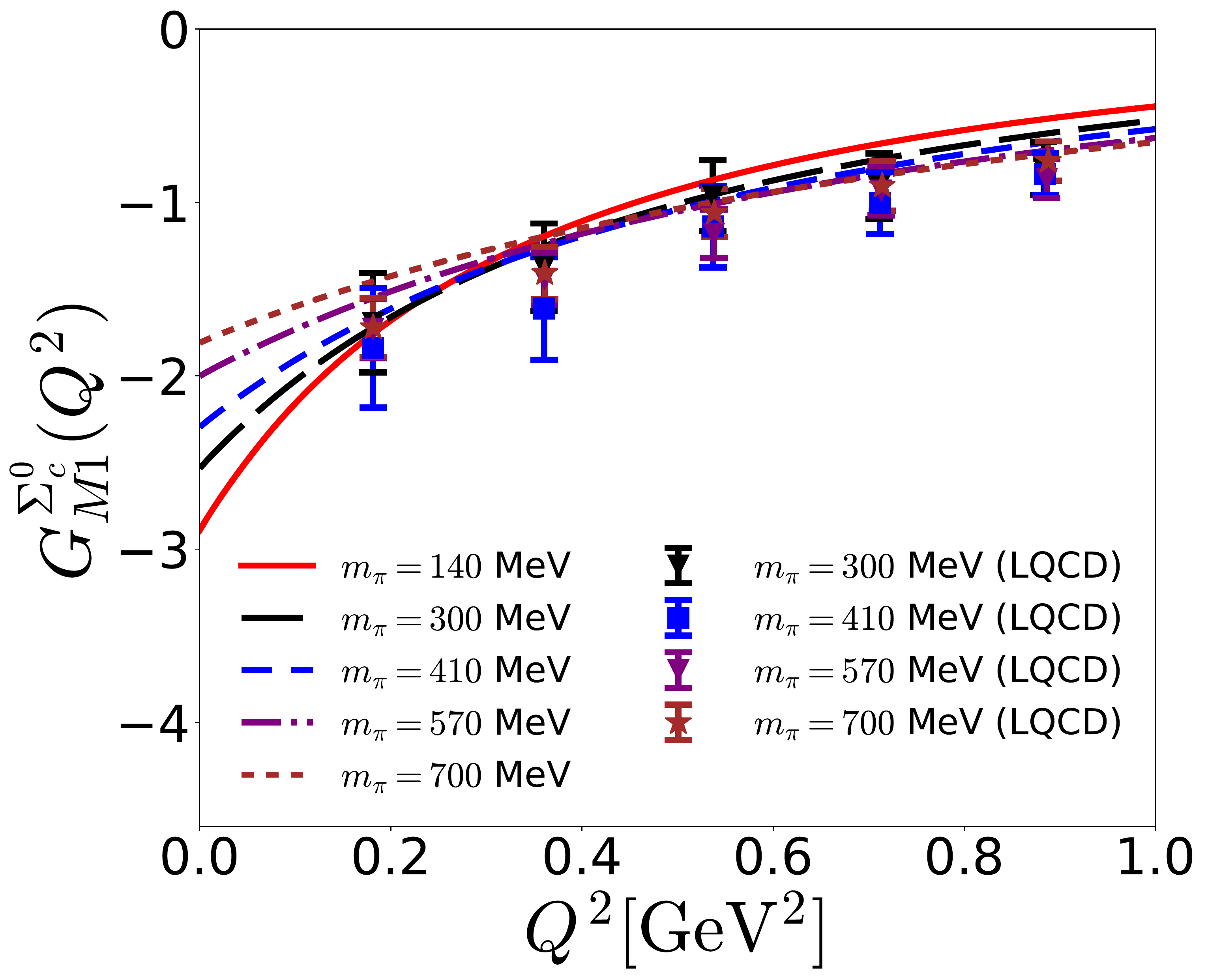}
\includegraphics[scale=0.28]{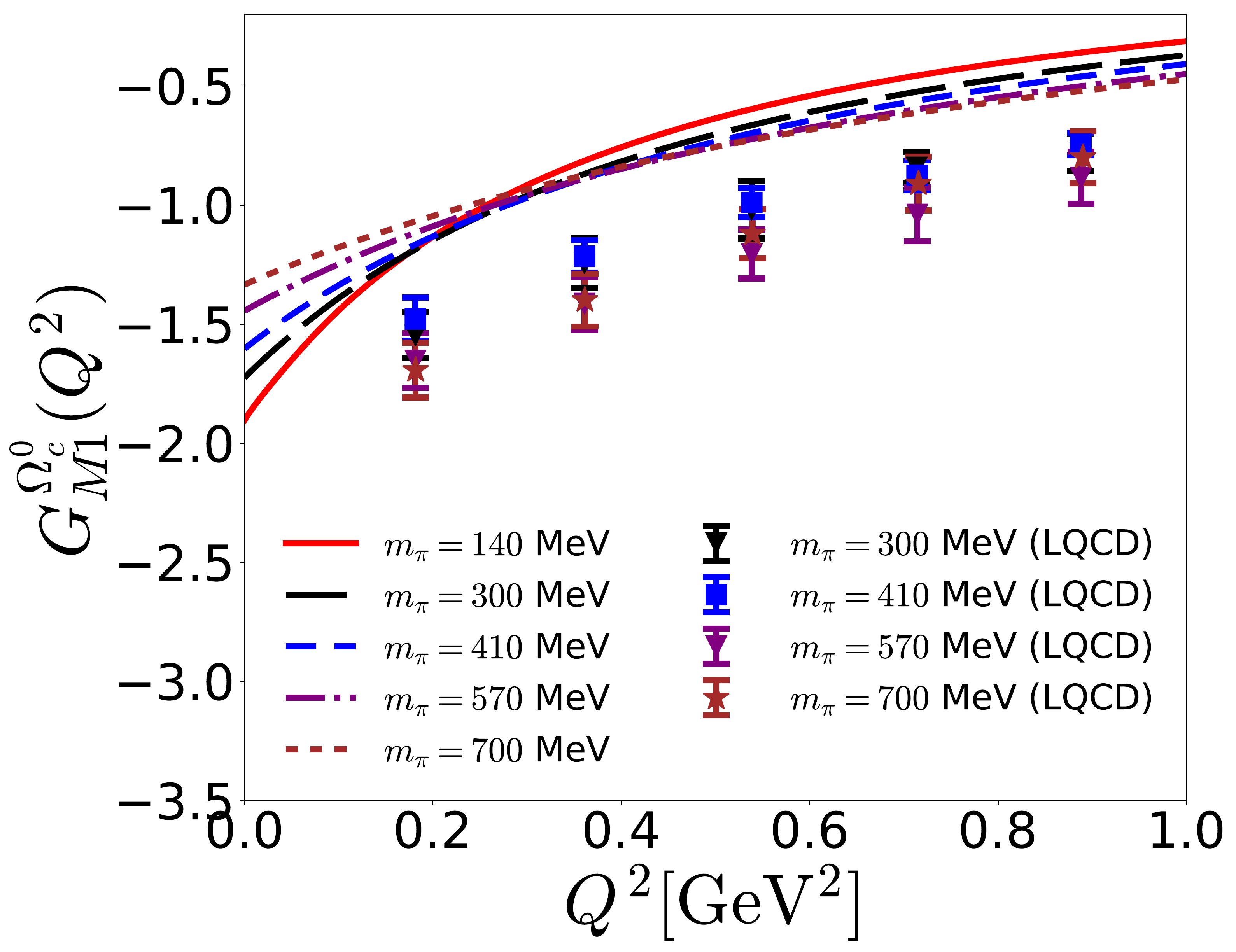}
\caption{Magnetic dipole form factors of the baryon sextet with
  spin 1/2 in comparison with the data from the lattice QCD. The data
  of the lattice QCD are taken from Ref.~\cite{Can:2013tna}.} 
\label{fig:4}
\end{figure}
Figure~\ref{fig:4} depicts the comparison of the present results for
the $M1$ form factors of the $\Sigma_c^{++}$, $\Sigma_c^{0}$, and
$\Omega_c^0$ heavy baryons with the corresponding lattice data. Note
that in order to compare the $Q^2$ dependence, we have normalized the
magnitudes of the magnetic form factors at $Q^2=0$ to be the same as
the lattice ones. In Ref.~\cite{Can:2013tna}, the chiral extrapolation
to the physical mass of the pion was performed. Here, we take the
values of the quadratic fitting obtained from
Ref.~\cite{Can:2013tna}:  $4.12$ for $\Sigma_c^{++}$, $3.80$ for
$\Sigma_c^0$, and $2.71$ for $\Omega_c$.  
The present results on the $Q^2$ dependence of the
$M1$ form factors are generally in qualitative agreement with the lattice
data. Again we find that the lattice results fall off more slowly,
compared to the present ones. In particular, the numerical results for
the $\Sigma_c^{++}$ and $\Sigma_c^0$ magnetic form factors get
closer to the lattice data as $m_\pi$ increases. That for $\Omega_c^0$
is also in line with the data.  

\begin{figure}[htp]
\centering
\includegraphics[scale=0.276]{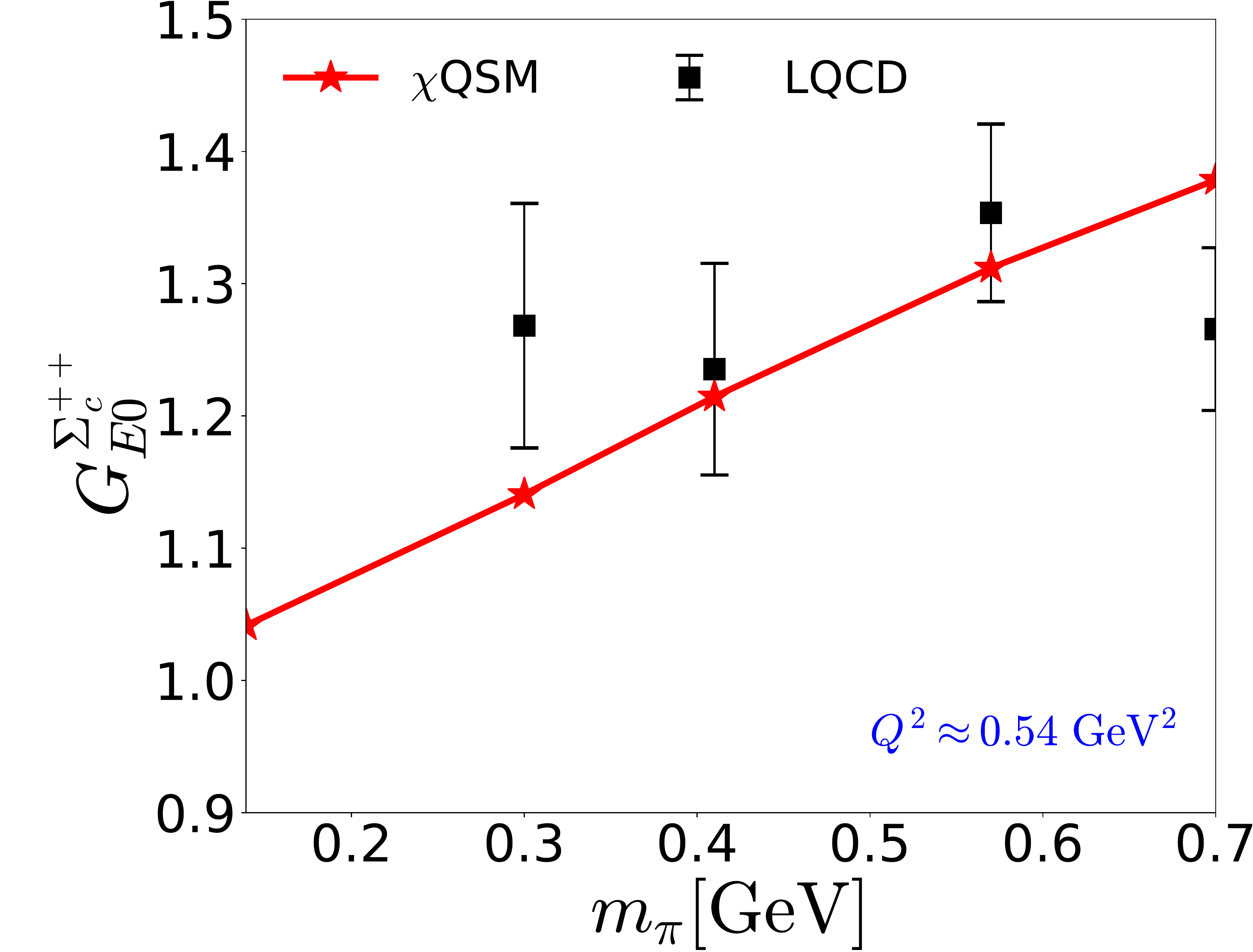}
\includegraphics[scale=0.276]{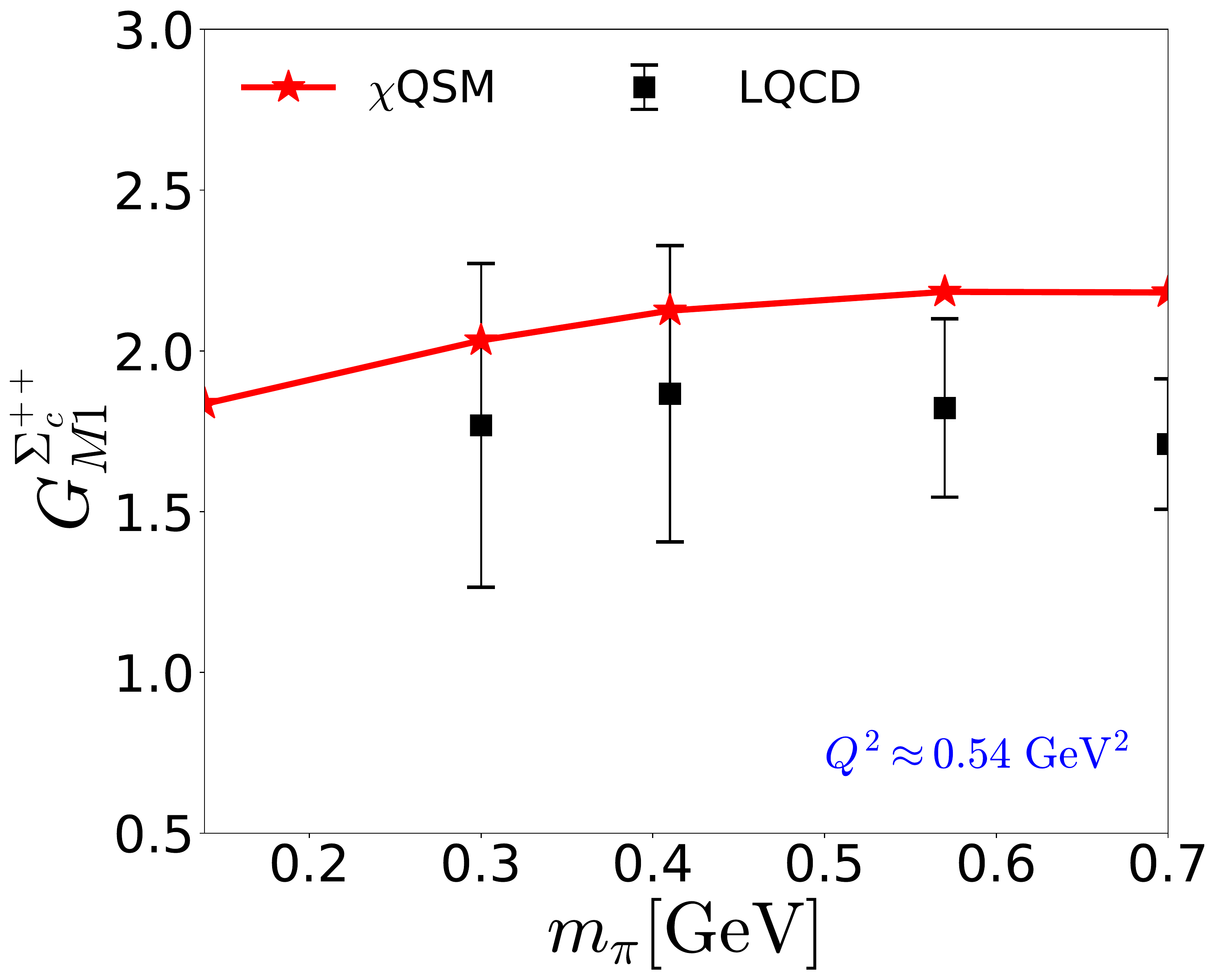}
\includegraphics[scale=0.276]{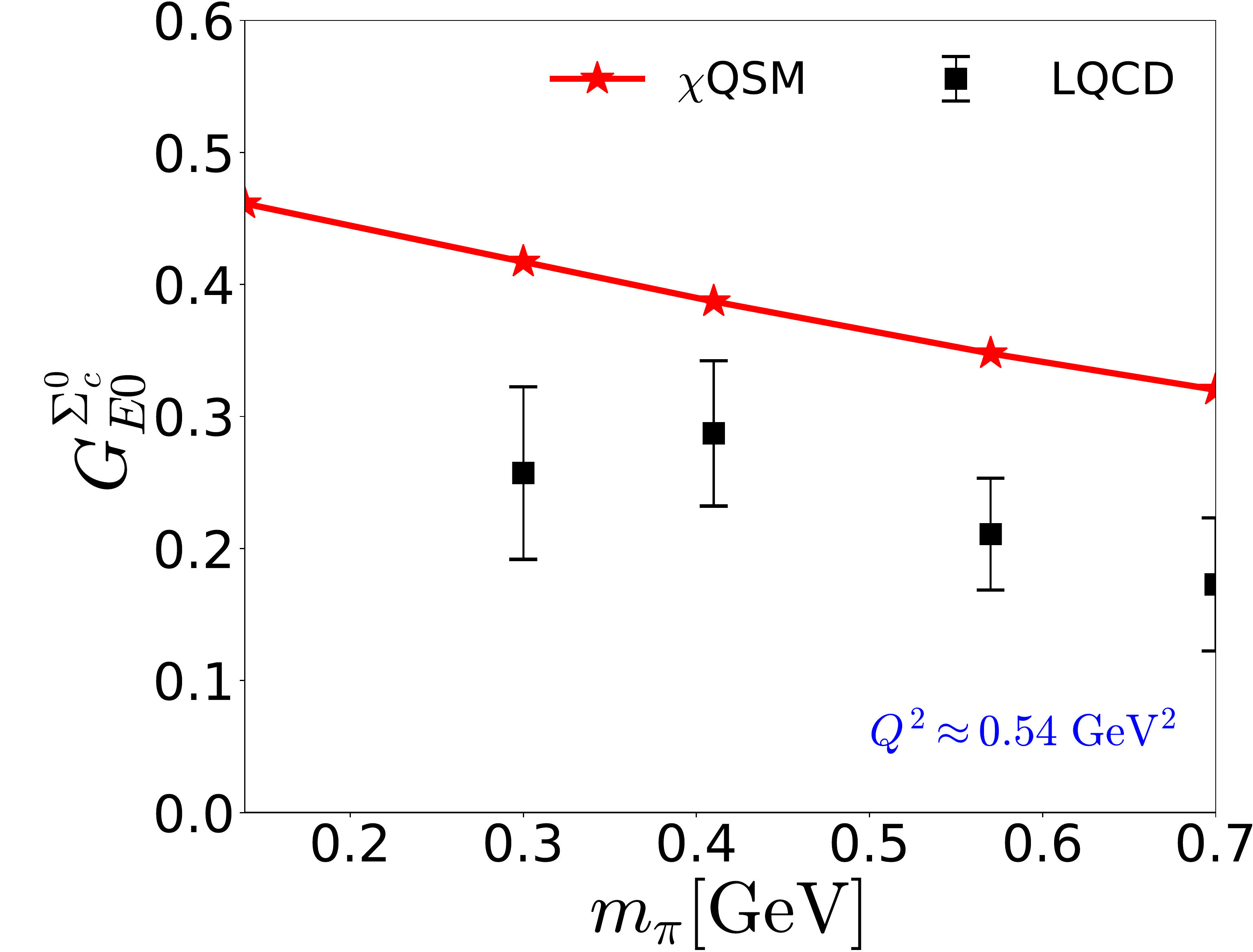}
\includegraphics[scale=0.276]{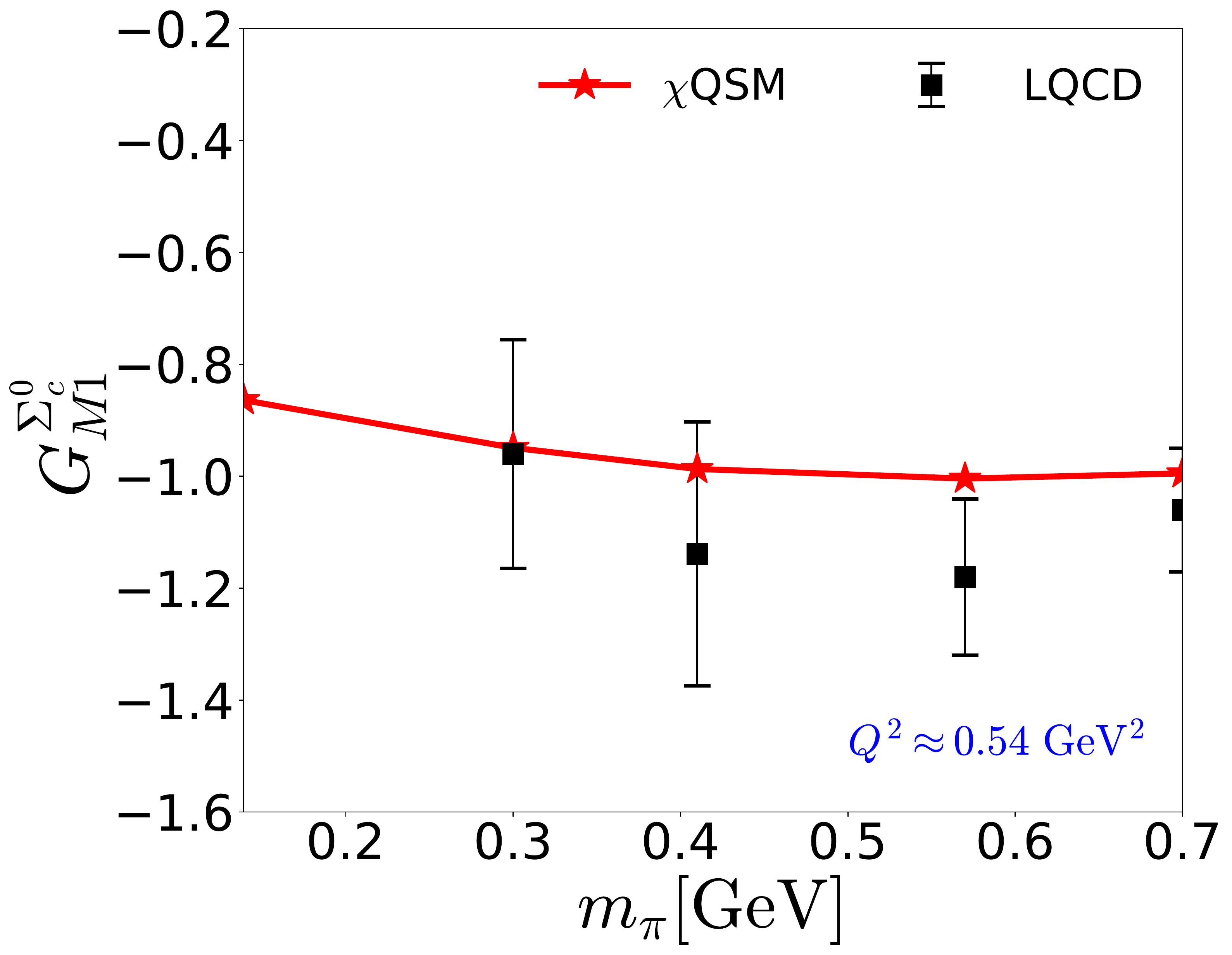}
\includegraphics[scale=0.276]{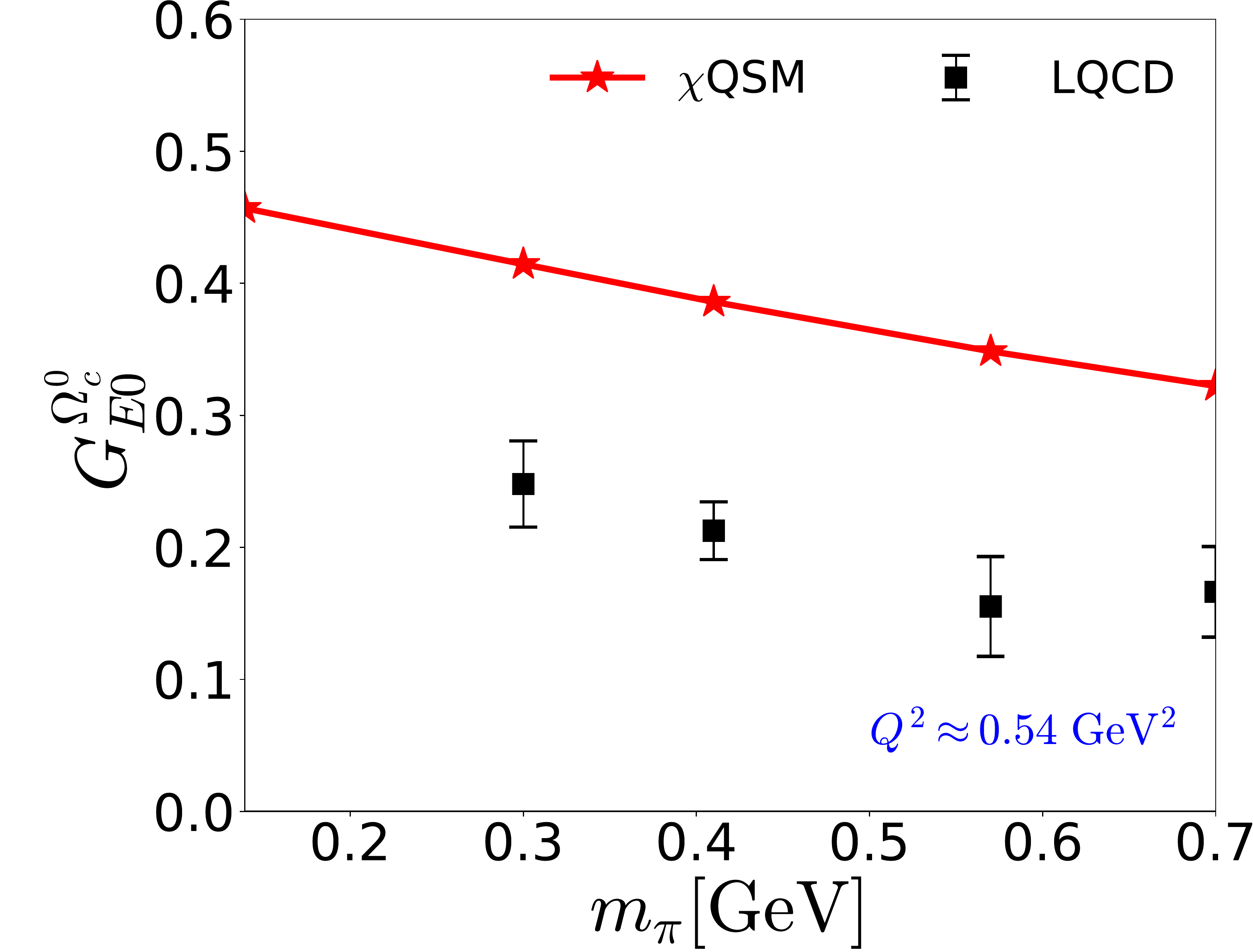}
\includegraphics[scale=0.276]{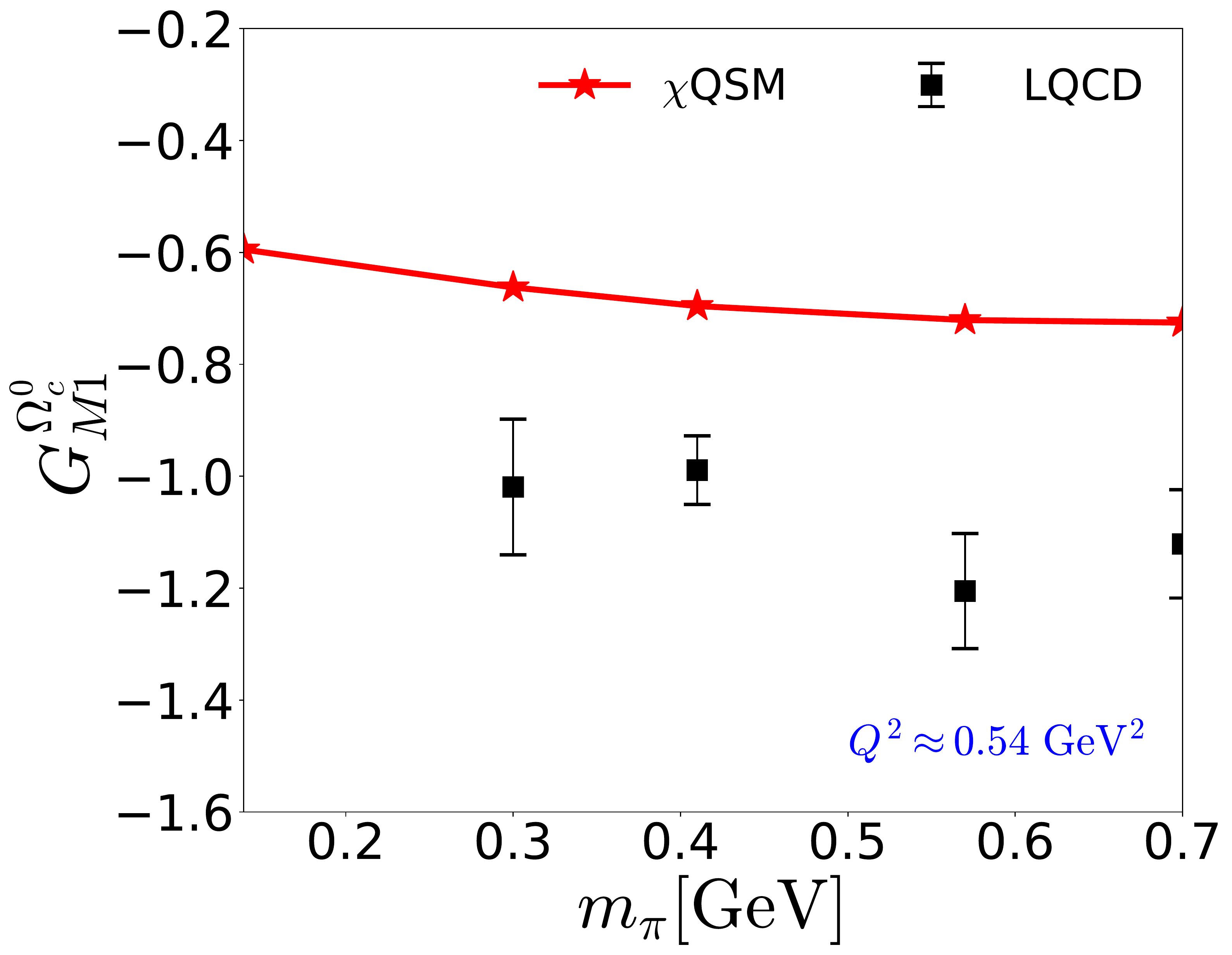}
\caption{Electromagnetic form factors as a function of the pion mass
  with the momentum transfer squared $Q^{2}\approx 0.54$~GeV fixed.
  The data of lattice QCD are taken from Ref.~\cite{Can:2013tna}.}  
\label{fig:5}
\end{figure}
Figure~\ref{fig:5} illustrates the results for the EM form factors of
the singly heavy baryons with spin 1/2 as functions of the pion mass,
with $Q^2$ fixed to be $0.54\,\mathrm{GeV}^2$. The numerical results
for the $\Sigma_c^{++}$ electric and magnetic form factors are in
agreement with the lattice data, as shown in the upper panel of 
Fig.~\ref{fig:5}. As for the EM form factors of the other heavy
baryons, the present results exhibit similar dependence on the pion
mass, compared with the lattice data. 

\section{Summary and conclusion}
In the present work, we have aimed at investigating the 
electromagnetic form factors of the lowest-lying singly
heavy baryons with spin 1/2 within the framework of the chiral
quark-soliton model, focusing on the comparison of the results with
recent lattice data. We first derived the profile functions of the
chiral soliton, employing the unphysical values of the pion mass. We
examined the limit of the heavy quark mass and showed that the
soliton mass consisting of the $N_c-1$ valence quarks converges on
$2m_0$. This implies that the pion mean fields get relatively suppressed as the
pion mass increases. Before we proceeded to compute the
electromagnetic form factors, we scrutinized the classical and
physical masses of the singly heavy baryons as the pion mass was
varied from $m_\pi^2=0.02\,\mathrm{GeV}^2$ to
$0.5\,\mathrm{GeV}^2$. The classical mass is in good agreement
with the lattice data on the $\Sigma_c$ and $\Omega_c$ masses. When we
considered the rotational $1/N_c$ corrections and the effects of
flavor SU(3) symmetry breaking, the present results for the $\Omega_c$
mass are in agreement with the lattice data. On the other
hand, those of the $\Sigma_c$ mass tends to rise faster than the data.  
We then calculated the electric form factors of the $\Sigma_c^{++}$,
$\Sigma_c^{0}$, and $\Omega_c^0$ for which there exsist the lattice
data. As the pion mass increases, the present results reproduce very
well the lattice data on the $\Sigma_c^{++}$ electric form factor. For
neutral heavy baryons, the results get closer to the lattice data.   
The results for the $\Sigma_c^{++}$ and $\Sigma_c^0$ magnetic form
factor are also in qualitative agreement with the lattice data. Those for the
$\Omega_c$ magnetic form factor show similar $Q^2$ dependence,
compared with the data. Finally, we compared the present results for
the electromagnetic form factors of the $\Sigma_c^{++}$, $\Sigma_c^0$,
and $\Omega_c^0$ as functions of the pion mass, fixing the momentum
transfer squared to be $Q^2=0.54\,\mathrm{GeV}^2$. Again, the results
for the $\Sigma_c^{++}$ and $\Sigma_c^0$ are in qualitative agreement with
the lattice data. The results for all other form factors are similar
dependence on the pion mass, compared with the lattice data.  

In conclusion, the present scheme describes well the
electromagnetic form factors of the lowest-lying singly heavy
baryons with spin 1/2, compared with those from lattice QCD. It indicates
that the singly heavy baryons with spin 1/2 are indeed well explained
in the pion mean-field approximation, i.e, in the chiral quark-soliton
model. The $1/m_Q$ corrections are expected to be marginal but are
very interesting issues, which will be considered in the near future.

\begin{acknowledgments}
The authors are grateful to Gh.-S. Yang for valuable discussions. 
They want to express the gratitude to K. U. Can for providing us with
the lattice data. The present work was supported by Inha University
Research Grant.  J.-Y. Kim is also supported by a DAAD doctoral
scholarship.   
\end{acknowledgments}

\appendix
\section{Fixing the model parameters \label{app:A}}
\begin{figure}
\centering
\includegraphics[scale=0.27]{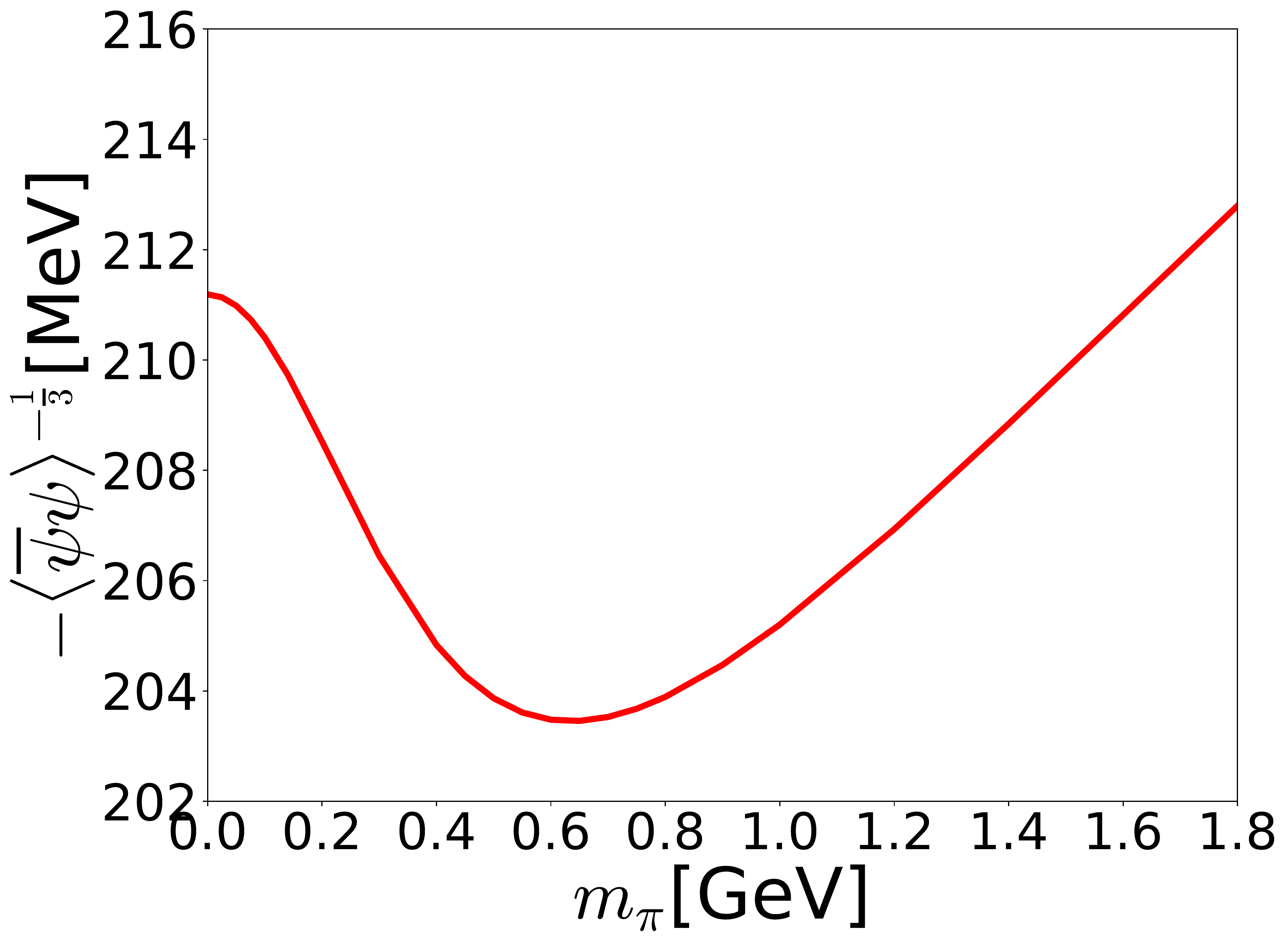}
\caption{Chiral condensate as a function of $m_{\pi}$ with fixed
  $f_{\pi}$ and $M$.} 
\label{fig:6}
\end{figure}

\begin{figure}
\centering
\includegraphics[scale=0.27]{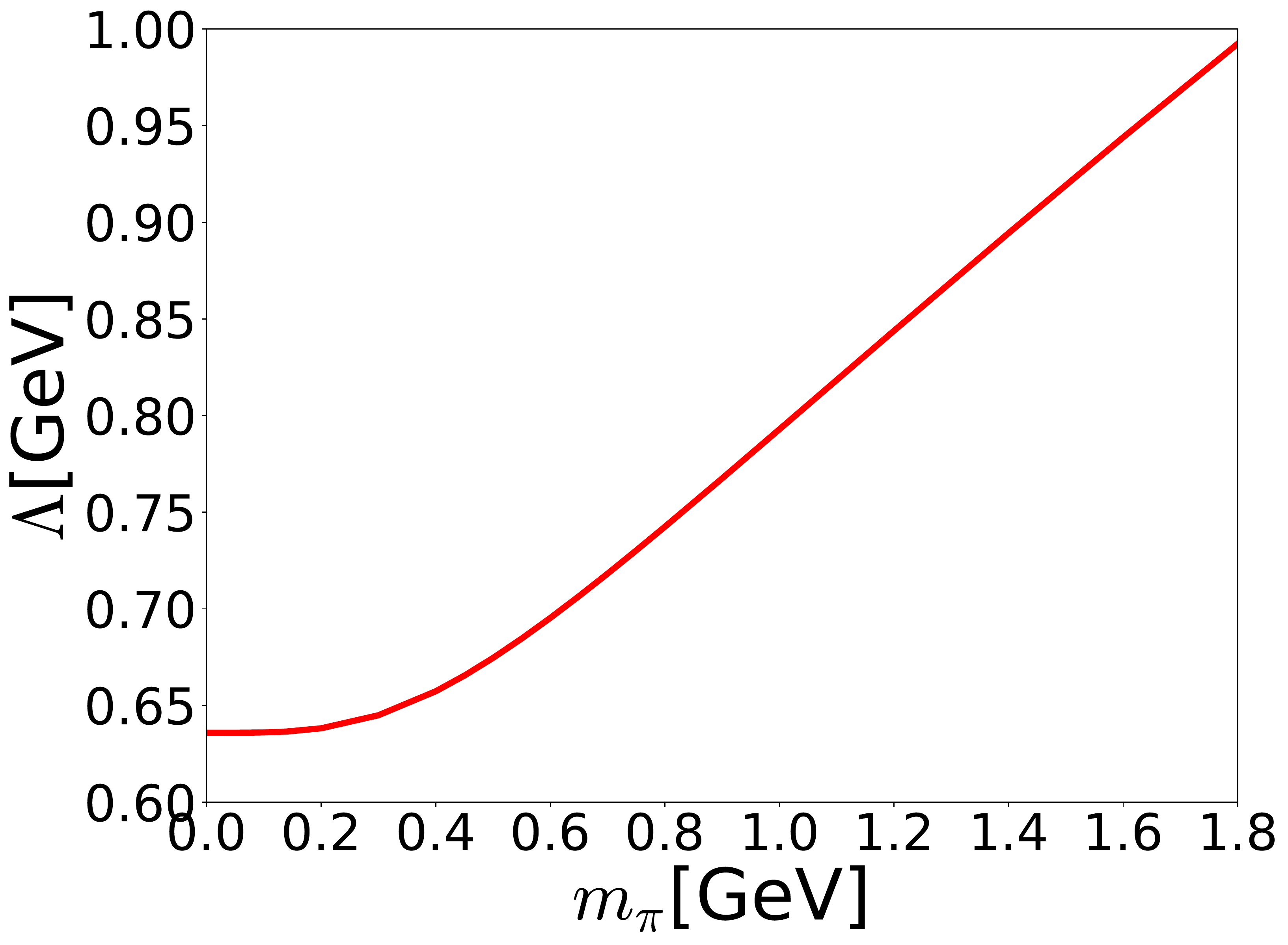}
\includegraphics[scale=0.27]{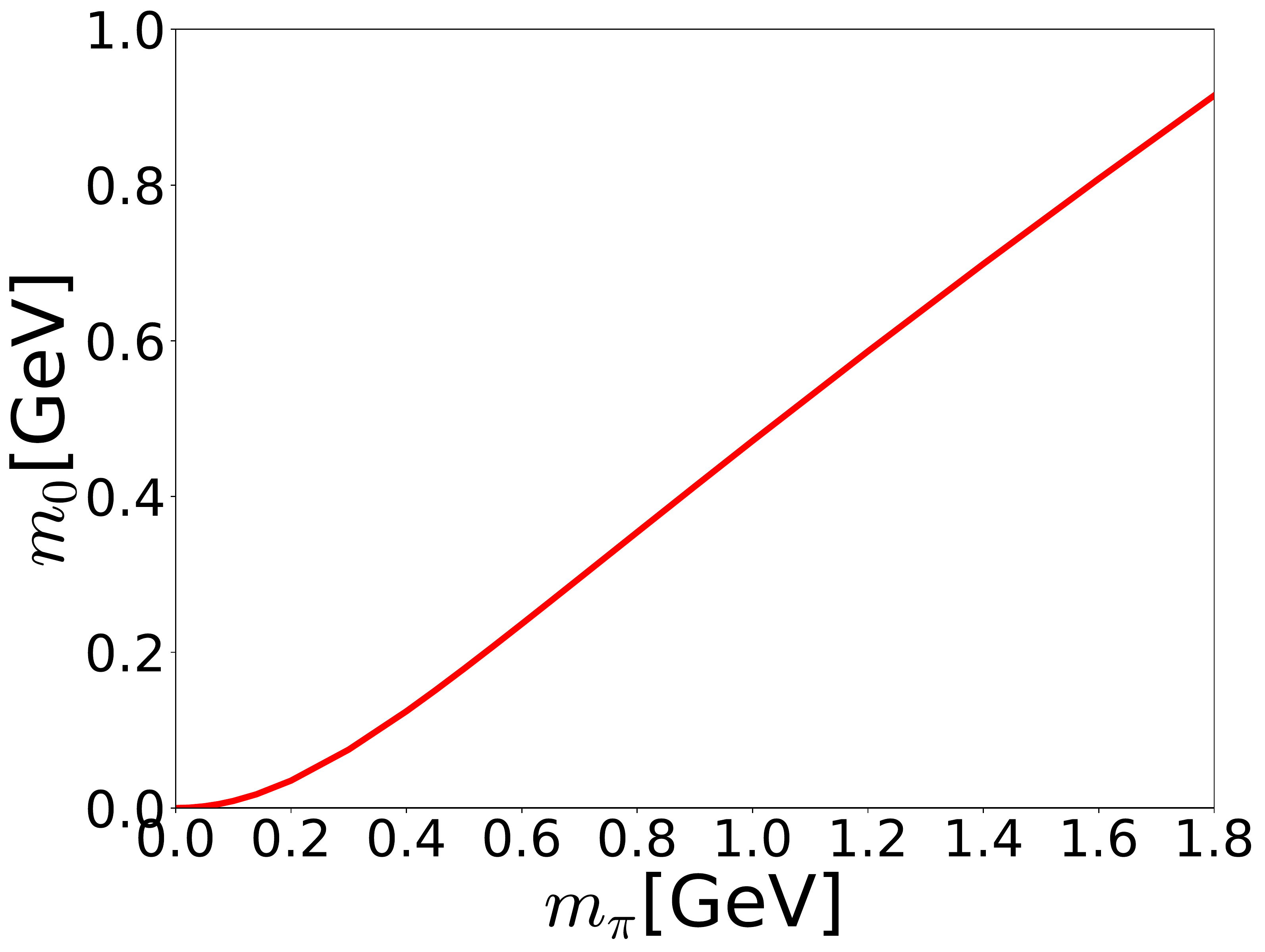}
\caption{Dependencies of parameters $\Lambda$ (left panel) and $m_{0}$
  (right panel) on $m_{\pi}$} 
\label{fig:7}
\end{figure}

Using the effective chiral action given in Eq.~\eqref{eq:echl}, one
can derive the expressions for the chiral condensate 
\begin{align}
\langle\overline{\psi}\psi\rangle  =
  -\int\frac{d^{4}p_{\mathrm{E}}}{(2\pi)^{4}}
  \frac{8N_{c}M'}{p^{2}_{\mathrm{E}}+M'^{2}} \bigg{|}_{reg} =
  -8N_{c}M'I_{1},
\label{eq:gap}
\end{align}
and for the pion decay constant 
\begin{align}
f^{2}_{\pi}= -\int\frac{d^{4}p_{\mathrm{E}}}{(2\pi)^{4}}
  \frac{4N_{c}M'^{2}}{(p^{2}_{\mathrm{E}}+M'^{2})^{2}} \bigg{|}_{reg}=
  8N_{c}M'^{2}I_{2}, 
\label{eq:pi}
\end{align}
where $M'=M+m_0$. $I_1$ and $I_2$ stand for the regularization
functions, which are expressed as 
\begin{align}
I_{1} &= \int^{\infty}_{\Lambda^{-2}} \frac{du}{u^{2}}
        \frac{e^{-uM'}}{(4\pi)^{2}}, \cr 
I_{2} &= \int^{\infty}_{\Lambda^{-2}} \frac{du}{2u}
        \frac{e^{-uM'}}{(4\pi)^{2}} \int^{1}_{0} d\beta
        e^{u\beta(1-\beta) m^{2}_{\pi}}. 
\end{align}
The pion  mass is determined by the pole position of the pion
propagator that is obtained by a low-energy effective chiral theory
given by Eq.~\eqref{eq:partftn} 
\begin{align}
m^{2}_{\pi} = \frac{m}{M}\frac{I_{1}}{I_{2}}.
\label{eq:pionmass}
\end{align}
When $m_\pi$ is the physical one, Eq.~\eqref{eq:pi} and
Eq.~\eqref{eq:pionmass} satisfy the Gell-Mann–Oakes–Renner(GOR)
relation  
\begin{align}
m^{2}_{\pi}f^{2}_{\pi} = - m_{0} \langle \overline{\psi} \psi \rangle
  + \mathcal{O}(m_{0}^{2}). 
\label{eq:GMOR}
\end{align}
Using the experimental value of the pion decay constant and fixing the
dynamical quark mass, we obtain the results for the 
current quark mass $m_{0}$, the cutoff mass $\Lambda$ and the chiral
condensate $\langle \overline{\psi} \psi \rangle$ as functions of
$m_\pi$. The results are drawn respectively in Fig~\ref{fig:6} and
Fig~\ref{fig:7}.  
In Fig.~\ref{fig:6}, we find that the results for the chiral
condensate decreases till the value of $m_\pi$ (or $m_0$) reaches $m_\pi\approx
0.6$ GeV and then increases monotonically as $m_{\pi}$ further
increases. Note that a lattice calculation~\cite{Durr:2013goa} predicts
monotonic increment of the chiral condensate as $m_\pi$
increases. Thus, while the present model gives a different behavior of
the chiral condensate with the lower values of $m_\pi$, it restores
the correct behavior when $m_\pi$ is larger than $m_\pi\approx 0.6$
GeV.  Figure~\ref{fig:7} shows how $\Lambda$ and $m_{0}$ depend on
$m_{\pi}$. Both the parameters increase as the pion mass
increases.


\end{document}